\documentclass[aps,amsmath,amssymb,prd,showpacs,floatfix,preprint,superscriptaddress,nofootinbib,11pt,a4paper]{article}
\usepackage{jheppub}
\usepackage{bm}
\usepackage{ulem}

\usepackage{axodraw4j}
\usepackage{pdfpages}
\usepackage{pstricks}
\usepackage{color}

\def\({\left(} \def\){\right)}

\allowdisplaybreaks[3]

\begin{document}

\title{The background field method and critical vector models}

\author[1]{Mikhail Goykhman,}
\author[2,3]{Vladimir Rosenhaus,}
\author[1]{and Michael Smolkin}

\affiliation[1]{The Racah Institute of Physics, The Hebrew University of Jerusalem, \\ Jerusalem 91904, Israel \\}
\affiliation[2]{School of Natural Sciences, Institute for Advanced Study, \\ 1 Einstein Drive, Princeton, NJ 08540, USA\\}
\affiliation[3]{Initiative for the Theoretical Sciences* \\ The Graduate Center, City University of New York\\
 365 Fifth Ave, New York, NY 10016, USA}

\abstract{
We use the background field method to systematically derive  CFT data for the critical $\phi^6$ vector model in three dimensions, and the Gross-Neveu model in dimensions $2\leq d \leq 4$. Specifically, we calculate the OPE coefficients and anomalous dimensions of various operators, up to next-to-leading order in the $1/N$ expansion.
\vfill
\noindent *On leave
}

\maketitle

\section{Introduction}

 Due to their theoretical and experimental importance,  conformal field theories (CFTs) have been extensively studied over the decades, through a broad range of techniques. One of the most prominent is the conformal bootstrap   \cite{Parisi:1972zm,Polyakov:1974gs,Ferrara:1973yt, ElShowk:2012ht, Simmons-Duffin:2016gjk}  which, rather than using a specific Lagrangian,  works with  general axiomatic parameters, such as the scaling dimensions of  local operators and their operator product expansion (OPE) coefficients. The focus of this paper is  inspired by the bootstrap program, but the CFT data is calculated using a complementary method: the diagrammatic $1/N$ expansion in the context of critical vector models \cite{Vasiliev:1981yc,Vasiliev:1981dg}.

The $1/N$ expansion is particularly useful when the theory is strongly coupled and the corresponding CFT cannot be studied perturbatively \cite{tHooft:1973alw,Witten:1979kh}. Thus, for instance, it provides an alternative to the classic $\epsilon$-expansion, originally formulated for the scalar field theory with a quartic interaction in $4-\epsilon$ dimensions \cite{Wilson:1971dc}.~\footnote{The Banks-Zaks fixed point is another example of a perturbatively tractable CFT \cite{Banks:1981nn,Bond:2018oco}.} Moreover, certain non-renormalizable quantum field theories which are not tractable within a traditional perturbative expansion become analytically controllable in the large $N$ framework. These include the most well-known vector models with a quartic interaction: the $\phi^4$  $O(N)$ vector model for bosons \cite{Parisi:1975im}, and the Gross-Neveu model for fermions \cite{Gross:1974jv,ZinnJustin:1991yn}. Another interesting example in this  class of models is the  cubic scalar vector model in $6-\epsilon$ dimensions \cite{Fisher:1978pf, Fei:2014yja,Fei:2014xta}.~\footnote{Another class of models which have been of interest recently are the SYK/tensor models \cite{SY, Kitaev, MS, GR4, Rosenhaus:2018dtp, Giombi:2017dtl, Klebanov:2018fzb}.}

A vector model which has been far less studied is the sextic $\phi^6$ $O(N)$ model, in three dimensions \cite{Pisarski:1982vz}, for recent studies see \cite{Litim:2016hlb,Juttner:2017cpr}. The model has  the notable feature -- which motivated our study of it --  of having a  UV conformal fixed point, at large $N$, directly in three dimensions. An $\epsilon$ expansion, as for the Wilson-Fisher fixed point in the quartic model, is not required. In this paper we compute various  anomalous dimensions and conformal three-point functions in the sextic $O(N)$ vector model, up to next-to-leading order in $1/N$. 

In the process of this study we found that, while the anomalous dimensions in critical vector models have been extensively studied, there has been much less study of the conformal three-point functions (the OPE data). 
We find that our method of computation -- in particular, applying the background field method \cite{Goldstone:1962es} in the context of  large $N$ conformal perturbation theory --  provides a simple and coherent framework. We apply this method to the standard vector models, extending known results in the literature.

In section~\ref{sec:preliminaries} we review the $\phi^6$ vector model, the location of the UV fixed point at large $N$, and the anomalous dimensions of the field. Then in section~\ref{sec:CFT data for the scalar sextic model} we compute several  conformal three-point functions at next-to-leading order in $1/N$. In section~\ref{sec:preliminaries gn} we review the Gross-Neveu model, and in section~\ref{sec: cft data in gross-neveu} we compute several  conformal three-point functions at next-to-leading order in $1/N$. In Appendix~\ref{app:vertex correction in ON vector model} we perform similar computations in the context of the quartic $O(N)$ vector model. In Appendix~\ref{B} we discuss the $\phi^6$ model, when  the field $\phi$ is integrated out.

\section{Review of the $\phi^6$ model}
\label{sec:preliminaries}

Consider the following three-dimensional $O(N)$ vector model with sextic interaction,
\begin{equation}
\label{3d original action}
S = \int d^3x \left(\frac{1}{2}\,(\partial\phi) ^ 2 + \frac{g_6}{6N^2}\, (\phi^2) ^ 3\right)\,.
\end{equation}
The beta function of $g_6$ vanishes identically in the large $N$ limit, and the model exhibits a line of fixed points, $0\leq g_6 \leq (4\pi)^2$, where the boundaries of the conformal window are found by demanding stability of the effective potential \cite{Bardeen:1983rv}, see also \cite{Rabinovici:1987tf,Marchais:2017jqc}. The $1/N$ corrections lift this degeneracy and only one isolated UV fixed point, which lies outside the stability region, survives at finite $N$ \cite{Pisarski:1982vz}.  The instability effect is, however, non-perturbative in $1/N$. Moreover, it is not clear whether it will persist at any finite $N$, see \cite{Fleming:2020qqx} and references therein for a recent discussion. In this section we introduce our notation, review the derivation of the UV fixed point, and calculate the anomalous dimensions of $\phi$ and $\phi^2$.

The model (\ref{3d original action}) can equivalently be  expressed in terms of the auxiliary fields $\sigma$ and $\rho$, with the action, 
\begin{equation}
\label{3d action in terms of aux fields}
S = \int d^3x \left(\frac{1}{2}\,(\partial\phi) ^ 2+ \frac{1}{\sqrt{N}} \, \sigma\,\phi^2 + \frac{g_6}{6\,\sqrt{N}}\,\rho^3 
-\,\sigma\rho\right)\,.
\end{equation}
The original action (\ref{3d original action}) is recovered from (\ref{3d action in terms of aux fields}) by first integrating over the
field $\sigma$ along the purely imaginary direction, which results in the delta-functional $\delta (\rho-\phi^2/\sqrt{N})$, and then integrating over $\rho$. Note that the fields $\rho$ and $\phi$ are the only dynamical degrees of freedom in the large $N$ limit, because the propagator of $\sigma$ is suppressed by $1/N$.

The propagator of the field $\phi$ in position space is,
\begin{equation}
\label{leading order phi propagator}
\langle\phi(x)\phi(0)\rangle = \frac{C_\phi}{|x|^{2\Delta_\phi}}\,,
\end{equation}
where the large $N$ amplitude and  scaling dimension are given by,
\begin{equation}
C_\phi = \frac{1}{4\pi}\,, \quad \Delta_\phi = \frac{1}{2}\,.
\end{equation}
The corresponding Feynman rule for the normalized field is,~\footnote{In our conventions, all propagators in the Feynman diagrams are normalized to unity, and therefore each Feynman graph should be multiplied by the corresponding amplitudes.} 
\begin{center}
  \begin{picture}(156,23) (31,-11)
  \thicklines
    \put(40,-2){\line(1,0){88}}
    \put(40,-2){\circle*{4}}
    \put(128,-2){\circle*{4}}
    \Text(78,0)[lb]{\scalebox{0.8}{$2\Delta_\phi$}}
    \Text(152,-15)[lb]{\scalebox{1.2}{$=\frac{1}{|x|^{2\Delta_\phi}}$}}
  \end{picture}
\end{center}
Notice that the quadratic part of the action (\ref{3d action in terms of aux fields}) is not diagonal, which leads to a non-trivial merging relation between the fields $\rho$ and $\sigma$,
\begin{center}
\begin{equation}
\label{rho and sigma correlator}
  \begin{picture}(244,23) (31,-26)
    \SetWidth{1.0}
    \SetColor{Black}
    \Line[](190,-7)(233,-7)
    \Vertex(190,-7){2}
    \Line[dash,dashsize=5](233,-7)(286,-7)
    \Vertex(286,-7){2}
    \Text(14,-13)[lb]{$\langle \rho(x_1)\sigma(x_2)\rangle ~= ~ - \,\delta^{(3)}(x_{12}) ~=~$}
  \end{picture}
\end{equation}
\end{center}
Here the non-dynamical field $\sigma$ is denoted by a dashed line.

\begin{figure}[t!]
\centering \noindent
\includegraphics[width=8cm]{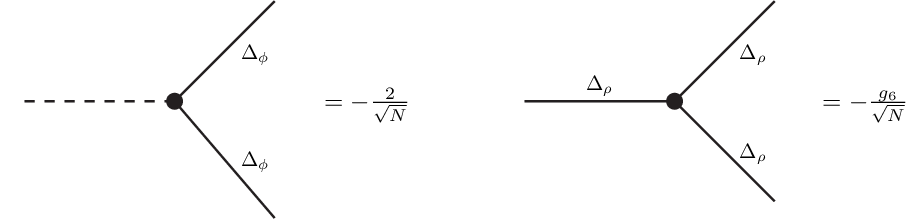}
\caption[]{Feynman rules for the cubic interactions in (\ref{3d action in terms of aux fields}).}
\label{fig:cubic interactions}
\end{figure}

The Feynman rules for the cubic interactions in (\ref{3d action in terms of aux fields}), $\phi^2\sigma$ and $\rho^3$,  are shown in Fig. \ref{fig:cubic interactions}. 
They generate kinetic terms for  $\sigma$ and $\rho$, respectively. However, the $\mathcal{O}(\rho^2)$ term in the effective action is suppressed by $1/N$ relative to the $\mathcal{O}(\sigma^2)$. Indeed, the $1/N$ factor from the cubic vertices is cancelled by $N$ fields $\phi$ running in the loop coming from $\sigma\phi^2$,
\begin{center}
  \begin{picture}(324,63) (27,0)
    \SetWidth{1.0}
    \SetColor{Black}
    \Vertex(264,34){4}
    \Arc[](292,34)(28,180,540)
    \Vertex(320,34){4}
      \Text(330,28)[lb]{$ \sigma(x_2) $}
    \Text(25,24)[lb]{$- S_\text{eff}\quad \supset \quad  C_\phi^2 \int d^3 x_1 \int d^3 x_2 {\sigma(x_1)\sigma(x_2)\over x_{12}^2}~ = ~ \sigma(x_1)$}
  \end{picture}
\end{center}
Inverting  the quadratic part of the effective action yields, 
\begin{equation}
\label{rho propagator}
\langle \rho(x)\rho(0)\rangle =  \frac{C_\rho}{|x|^{2\Delta_\rho}}\,, 
\end{equation}
where the large $N$ amplitude and  scaling dimension of the field $\rho$ are given by,
\begin{equation}
\label{leading N rho params}
 C_\rho = 2 C_\phi^2=\frac{1}{8\pi^2}~, \ \ \ \ \ \ \ \ \Delta_\rho = 1\,.
\end{equation}
The corresponding Feynman rule for the normalized field is,
\begin{center}
  \begin{picture}(156,23) (31,-11)
    \SetWidth{1.0}
    \SetColor{Black}
    \Line[](40,-2)(128,-2)
    \Vertex(40,-2){2}
    \Vertex(128,-2){2}
    \Text(78,3)[lb]{\scalebox{1}{$2\Delta_\rho$}}
    \Text(152,-13)[lb]{\scalebox{1.2}{$=\frac{1}{|x|^{2\Delta_\rho}}~.$}}
  \end{picture}
\end{center}

In the next subsection, we are going to use the background field method to calculate the effective cubic vertex in the effective action, at  next-to-leading order in the $1/N$ expansion. In preparation for this, here we establish an additional Feynman rule in the presence of a background field $\bar\rho(x)$, {\it i.e.,}  we substitute $\rho\to\rho+\bar \rho$ into (\ref{3d action in terms of aux fields}) to get an action in the presence of $\bar\rho(x)$. The relevant Feynman rule is based on the observation that the 
linear coupling $\rho\,\sigma$ becomes $(\rho + \bar\rho)\sigma$. Hence, $\bar\rho$ is a source field for $\sigma$. In particular, it induces a background field $\bar\sigma(x)$, which can be derived by carrying out the gaussian integral over $\phi$ and eliminating the term $ \bar\rho\sigma$ by a field redefinition $\sigma\to \sigma + \bar\sigma$, where
%
\begin{center}
\begin{equation}
\label{sigma background}
  \begin{picture}(180,35) (39,35)
    \SetWidth{1.0}
    \SetColor{Black}
    \Vertex(180,59){4}
    \Line[](180,59)(252,59)
    \Vertex(252,59){2}
    \Text(215,63)[lb]{\scalebox{1}{$4$}}
    \Text(170,67)[lb]{\scalebox{1}{$ \bar \rho(x_1)$}}
    \Text(255,67)[lb]{\scalebox{1}{$x$}}
    \Text(-50,48)[lb]{\scalebox{1.2}{$\bar\sigma(x)={4 \over  \pi^2} {}\int d^3 x_1 {\bar\rho(x_1)\over |x-x_1|^4} \quad = \quad {4\over \pi^2}~\times$}}
  \end{picture}
 \end{equation}
\end{center}
A quick consistency check can be carried out by contracting both sides of this expression with $\bar\rho(x_2)$,
and using (\ref{rho and sigma correlator}), (\ref{rho propagator}), (\ref{leading N rho params}), and the inverse propagator relation ($d=3$, $\Delta = 2$),
\begin{equation}
\label{inverse propagator relation}
\frac{1}{\pi^d}\,\frac{\Gamma(d-\Delta)\Gamma(\Delta)}{\Gamma\left(\frac{d}{2}-\Delta\right)
\Gamma\left(\Delta-\frac{d}{2}\right)}\,\int d^dx_2\,
\frac{1}{|x_{12}|^{2\Delta}|x_{23}|^{2(d-\Delta)}} = \delta^{(d)}(x_1-x_3)\,.
\end{equation}

There are simple diagrammatical rules for performing some of the integrals involved in the conformal perturbation theory calculations we will encounter. For instance, a simple loop diagram satisfies additivity (in position space), 
\begin{center}
  \begin{picture}(197,71) (39,-5)
    \SetWidth{1.0}
    \SetColor{Black}
    \Vertex(40,31){2}
    \Arc[clock](82.5,-4)(55.057,141,39)
    \Vertex(125,31){2}
    \Arc[clock](82.5,65)(55.057,-39,-141)
    \Text(135,28)[lb]{$=$}
    \Vertex(155,31){2}
    \Line[](155,31)(215,31)
    \Vertex(215,31){2}
    \Text(165,35)[lb]{\scalebox{0.8}{$2\Delta_1+2\Delta_2$}}
    \Text(77,55)[lb]{\scalebox{0.8}{$2\Delta_1$}}
    \Text(80,-3)[lb]{\scalebox{0.8}{$2\Delta_2$}}
  \end{picture}
\end{center}
In addition, there is a propagator merging relation of the form,
\begin{equation}
\label{propagator merging relation}
 \int d^d x_2 ~ {1\over |x_{12}|^{2\Delta_1} |x_{23}|^{2\Delta_2}}= {U(\Delta_1,\Delta_2,d-\Delta_1-\Delta_2)\over |x_{13}|^{2(\Delta_1+\Delta_2)-d}} ~,
\end{equation}
where 
\begin{equation}
U(\Delta_1,\Delta_2,\Delta_3) = \pi^\frac{d}{2}\,\frac{\Gamma\left(\frac{d}{2}-\Delta_1\right)\Gamma\left(\frac{d}{2}-\Delta_2\right)\Gamma\left(\frac{d}{2}-\Delta_3\right)}
{\Gamma(\Delta_1)\Gamma(\Delta_2)\Gamma(\Delta_3)}\,.
\end{equation}
This relation can be represented diagrammatically as,~\footnote{The middle vertex on the left-hand side is integrated over.}
\begin{center}
  \begin{picture}(271,37) (49,-10)
    \SetWidth{1.0}
    \SetColor{Black}
    \Vertex(30,7){2}
    \Line[](30,7)(135,7)
    \Vertex(135,7){2}
    \Vertex(85,7){4}
    \Text(52,12)[lb]{\scalebox{0.8}{$2\Delta_1$}}
    \Text(107,12)[lb]{\scalebox{0.8}{$2\Delta_2$}}
    \Text(151,4)[lb]{$=$}
    \Vertex(175,7){2}
    \Line[](175,7)(260,7)
    \Vertex(260,7){2}
    \Text(190,12)[lb]{\scalebox{0.8}{$2(\Delta_1+\Delta_2)-d$}}
    \Text(270,2)[lb]{\scalebox{1}{$\times~ U(\Delta_1,\Delta_2,d{-}\Delta_1{-}\Delta_2)$}}
  \end{picture}
\end{center}
Finally, the star-triangle relation for the cubic scalar vertex is given by,
\begin{equation}
\label{uniqueness scalar}
\int d^dx_4\,\frac{1}{|x_{14}|^{2\Delta_1}|x_{24}|^{2\Delta_2}|x_{34}|^{2\Delta_3}}
= \frac{U (\Delta_1,\Delta_2,\Delta_3)}{|x_{12}|^{d-2\Delta_3}
|x_{13}|^{d-2\Delta_2}|x_{23}|^{d-2\Delta_1}} ~ , \quad \Delta_1+\Delta_2+\Delta_3 = d ~.
\end{equation}
Diagrammatically it takes the form,
\begin{center}
\scalebox{0.8}{
  \begin{picture}(390,106) (29,-27)
    \SetWidth{1.0}
    \SetColor{Black}
    \Line(96,28)(144,76)
    \Line(144,-26)(96,28)
    \Line[](96,28)(30,28)
    \Vertex(96,28){4}
    \Vertex(144,76){2}
    \Vertex(144,-26){2}
    \Vertex(30,28){2}
    \Text(160,22)[lb]{\scalebox{1}{$= U(\Delta_1,\Delta_2,\Delta_3)~\times$}}
    \Line(266,28)(380,76)
    \Line(380,-26)(266,28)
    \Line(380,76)(380,-26)
    \Text(60,32)[lb]{\scalebox{1}{$2\Delta_1$}}
    \Text(103,58)[lb]{\scalebox{1}{$2\Delta_2$}}
    \Text(103,-8)[lb]{\scalebox{1}{$2\Delta_3$}}
    \Text(303,62)[lb]{\scalebox{1}{$d-2\Delta_3$}}
    \Text(303,-18)[lb]{\scalebox{1}{$d-2\Delta_2$}}
    \Text(385,22)[lb]{\scalebox{1}{$d-2\Delta_1$}}
  \end{picture}
  }
\end{center}

\subsection{UV fixed point}
\label{sec: UV fixed point}

\begin{figure}[t!]
\centering \noindent
\includegraphics[width=12cm]{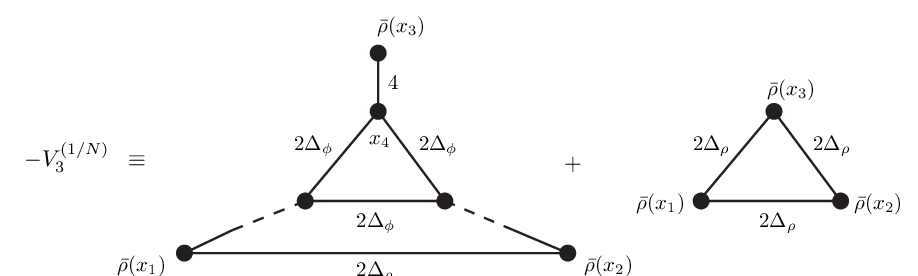}
\caption[]{Subleading correction to the cubic vertex of the effective action. $\bar\rho(x)$ is a fixed background field.}
\label{fig:phi6}
\end{figure}

The beta function of $g_6$ vanishes in the large $N$ limit. However, the two diagrams shown in Fig.~\ref{fig:phi6} induce a non-trivial RG flow at the next-to-leading order.  These diagrams represent a $1/N$ correction to the cubic vertex of the effective action in the presence of the background field $\bar\rho(x)$,\footnote{When calculating the first diagram in Fig.~\ref{fig:phi6}, we used expression (\ref{sigma background}) for the induced background field $\bar\sigma(x)$. The latter is coupled to $\phi^2$ via a cubic vertex with  amplitude $-2/\sqrt{N}$. When writing down the contributions of these diagrams to $V_3^{(1/N)}$, we accounted for the symmetry factors $1/2!$ and $1/3!$, respectively.}
\begin{eqnarray}
V_3^{(1/N)}&=& {16 \, C_\phi^3 \, C_\rho \over \pi^2\sqrt{N}}  \(\frac{g_6}{ \sqrt{N}}\)^2 
    \int  \prod_{i=1}^{3} d^3x_i  \int d^3 x_4 ~  {\bar\rho(x_1)\bar\rho(x_2) \bar\rho(x_3) \over |x_{12}|^3 |x_{14}| |x_{24}||x_{34}|^4} 
  \nonumber\\
 &+& {C_\rho^3 \over 6}  \({g_6\over \sqrt{N}}\)^3 
   \int  \prod_{i=1}^{3} d^3x_i  ~ {\bar\rho(x_1)\bar\rho(x_2) \bar\rho(x_3) \over |x_{12}|^2 |x_{13}|^2 |x_{32}|^2}
  ~.
  \label{eff6vertex}\
\end{eqnarray}
The integral over $x_4$ in the first line can be carried out through the use of the star-triangle relation (\ref{uniqueness scalar}).~\footnote{The integral actually suffers from a power law divergence. However, such divergences depend on the choice of regularization scheme. For instance, they are absent within dimensional regularization. Therefore, we ignore them in what follows. Indeed, a similar remark applies to the integral identities, such as the propagator merging relation and the star-triangle relation. On the other hand, in general these relations are singular if the integral has a logarithmic divergence.} We get,
\begin{equation}
\label{V3 1 over N}
V_3^{(1/N)}
= {- g_6^2\over 16\pi^6 N^{3/2}}\Big( 1-{g_6\over 192} \Big)  \int  \prod_{i=1}^{3} d^3x_i ~ {\bar\rho(x_1)\bar\rho(x_2) \bar\rho(x_3) \over |x_{12}|^2 |x_{13}|^2 |x_{23}|^2}~.
\end{equation}

Naively, this vertex is conformally invariant, regardless of the value of $g_6$. However, this is deceptive, since it exhibits a logarithmic UV divergence when the three fields  $\bar\rho(x)$ collide.  This  results in a non-trivial RG flow for $g_6$. To derive the associated beta function, it is sufficient to set all the fields to be at the same point $x_1$.  Using the propagator merging relation (\ref{propagator merging relation}) to integrate over $x_2$, and introducing both a  UV cut-off $\mu_0$ and an IR cut-off $\mu$ to regularize the integral over $x_3$ yields,\footnote{We use UV/IR terminology in the Wilsonian sense. More specifically, the coupling $g_6$ of the original action (\ref{3d action in terms of aux fields}) is
defined at a certain (UV) scale $\mu_0$. It is denoted by $g_6(\mu_0)$
in (\ref{V3 1 over N divergence}) and (\ref{g6 flow}). We are interested in finding the value of
$g_6(\mu)$ at a lower (IR) scale  $\mu$. If the theory is conformal,
it looks the same at all scales, and there is no difference between
the couplings. Unless $g_6=192$, this is not the case in the $\phi^6$ model.

Indeed, the $\phi^6 \sim \rho^3$ interaction in (\ref{3d action in terms of aux fields}) accumulates correction (\ref{V3 1 over N divergence})
at a lower (IR) scale $\mu$. This correction vanishes when
$\mu=\mu_0$. Setting
$\bar\rho$ to a constant value (zero mode) in (\ref{V3 1 over N}) and evaluating the integrals over a
shell $\mu_0^{-1}\lesssim |x| \lesssim \mu^{-1}$ is an alternative way of deriving (\ref{V3 1 over N divergence}). This is enough to get the RG flow of $g_6$, because the sextic
interaction contributes to the effective potential.}
\begin{equation}
 \label{V3 1 over N divergence}
V_3^{(1/N)}= - {g_6^2(\mu_0)\over 4\pi^2 N^{3/2}}\Big( 1-{g_6(\mu_0)\over 192} \Big)
 \log(\mu_0/\mu) \int d^3x_1\, \bar\rho^3(x_1) + \ldots
  ~,
\end{equation}
where ellipsis encode finite terms.

Combining this result with the leading order cubic vertex, we deduce the following RG flow equation for the  coupling, 
\begin{equation}
\label{g6 flow}
g_6(\mu)=g_6(\mu_0) - {3 g_6^2(\mu_0)\over 2\pi^2 N}\Big( 1-{g_6(\mu_0)\over 192} \Big) \log(\mu_0/\mu) 
  ~.
\end{equation}
Equivalently, the beta function is, 
\begin{equation}
 \beta(g_6)= \mu {d g_6\over d\mu} =  {3 g_6^2\over 2\pi^2 N} \(1-{g_6\over 192}\) + \mathcal{O}(1/N^2)~.
 \label{beta}
\end{equation}
In particular, the model exhibits a UV stable fixed point at $g_6^*=192$, as originally found in \cite{Pisarski:1982vz}. 
Note also that the $1/N$ correction, $V_3^{(1/N)}$, to the cubic vertex vanishes at the UV fixed point. We will use this fact later in section \ref{OPE}, when we derive the conformal three-point function $\langle \rho \rho \rho\rangle$.

\subsection{Anomalous dimensions of $\phi$ and $\phi^2$}
\label{known anomalous dims}

In this section we apply the background field method \cite{Goldstone:1962es} to derive the anomalous dimension of $\phi$ and of $\rho\sim \phi^2$. \footnote{The identification $\rho\sim\phi^2$ follows from the equation of motion operator, {\it i.e.}, vary (\ref{3d action in terms of aux fields}) with respect to $\sigma$.} These two operators are special because, unlike composite operators with higher powers of $\phi$, their anomalous dimensions vanish at leading and next-to-leading order in $1/N$. One has to consider ${\cal O}(1/N^2)$ diagrams to find the first non-trivial correction.

\begin{figure}[t!]
\centering \noindent
\includegraphics[width=9cm]{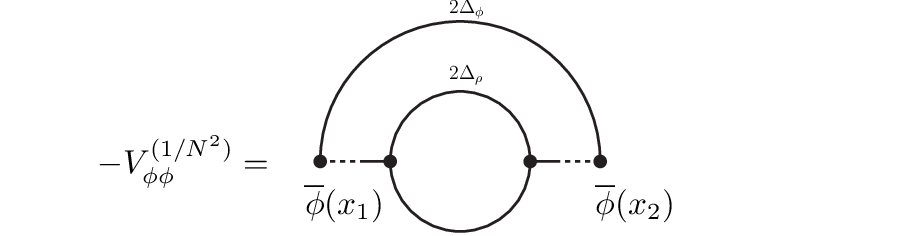}
\caption[]{The $\mathcal{O}(1/N^2)$ correction to the quadratic effective action of $\phi$, where $\bar\phi(x)$ is a given background field.}
\label{fig:Vphiphi}
\end{figure}

For instance, the anomalous dimension of $\phi$ (denoted by $\gamma_\phi$) is entirely fixed by the diagram in Fig. \ref{fig:Vphiphi}. It represents the leading order correction to the quadratic effective action of the scalar field.
Using the Feynman rules gives,
\begin{equation}
- V_{\phi\phi}^{(1/N^2)}=  {1\over 4} \Big({g_6\over \sqrt{N}} \Big)^2\Big( {2\over \sqrt{N}} \Big)^2 C_\phi C_\rho^2 
 \int d^3 x_1 \int d^3 x_2 {\bar\phi(x_1)\bar\phi(x_2)\over  |x_{12}|^5} ~.
\end{equation}
This term is singular in the vicinity of $x_1\sim x_2$. The logarithmic divergence is associated with the anomalous dimension of $\phi$, and we isolate it by expanding the background field $\bar\phi(x_2)$ around $x_1$,
\begin{equation}
V_{\phi\phi}^{(1/N^2)}= - {g_6^2\over 2 N^2} C_\phi C_\rho^2 
 \int d^3 x_1 ~ \bar\phi(x_1) \partial_\mu \partial_\nu \bar\phi(x_1) \int d^3 x_2 {x_{12}^\mu x_{12}^\nu\over  |x_{12}|^5} + \ldots~,
\end{equation}
where ellipsis encode terms which do not contribute to $\gamma_\phi$.  Introducing a spherical sharp cut-off $\mu$, and substituting the fixed point value $g_6^*=192$, yields 
\begin{equation}
V_{\phi\phi}^{(1/N^2)}=  {96\over \pi^4 N^2} 
 \int d^3 x ~ \partial_\mu \bar\phi \partial^\mu \bar\phi~ \log(\mu|x|)+ \ldots~.
\end{equation}
The value of $\gamma_\phi$ can be obtained by comparing this expression with the full propagator of $\phi$,
\begin{equation}
\label{corrected phi propagator}
\langle \phi(x)\phi(0)\rangle =\mu^{-2\gamma_\phi}\frac{C_\phi\,(1+A_\phi)}{|x|^{1+2\gamma_\phi}}\,,
\end{equation}
where $A_\phi$ is associated with higher order corrections to the leading order amplitude $C_\phi$. Hence, 
\begin{equation}
 \gamma_\phi= {96\over \pi^4 N^2} + \mathcal{O}(1/N^3) ~.
\end{equation}
in full agreement with \cite{Pisarski:1982vz}.

Next we consider the field $\rho$. Its full propagator can be written as,
\begin{equation}
\label{corrected rho propagator}
\langle \rho(x)\rho(0)\rangle =\mu^{-2\gamma_\rho}\frac{C_\rho\,(1+A_\rho)}{|x|^{2(1+\gamma_\rho)}}\,,
\end{equation}
where $\gamma_\rho$ is the anomalous dimension, and $A_\rho$ is associated with $1/N$ corrections to the leading order amplitude $C_\rho$. 
\begin{figure}[t!]
\centering \noindent
\includegraphics[width=12cm]{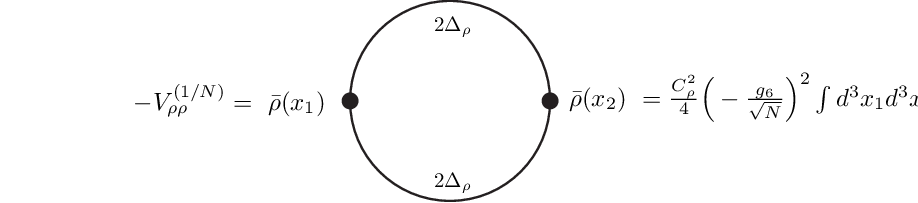}
\caption[]{The $\mathcal{O}(1/N)$ correction to the quadratic effective action of $\rho$, where $\bar\rho$ is the background field. The solid dots represent the cubic vertex with one of the legs substituted by the background $\bar\rho$, {\it i.e.,} $-{g_6\over \sqrt{N}} \bar\rho$, and $1/4$ is the symmetry factor.}
\label{fig:Vrhorho1}
\end{figure}
 
\begin{figure}[t!]
\centering \noindent
\includegraphics[width=13cm]{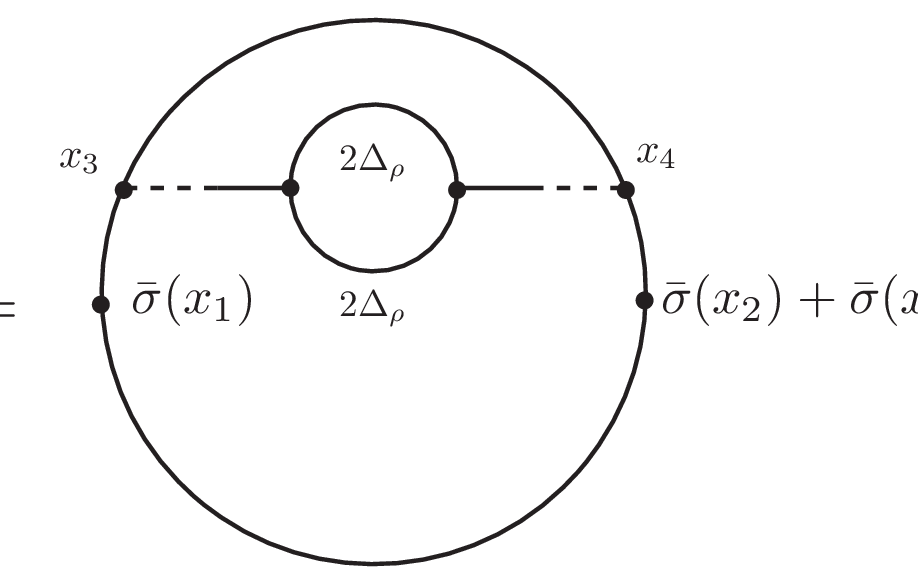}
\caption[]{The $\mathcal{O}(1/N^2)$ correction to the quadratic effective action of $\rho$. The background field is $\bar\sigma\sim\bar\rho$, given by (\ref{sigma background}). All propagators on the outer circles of these diagrams correspond to the field $\phi$ (the label $2\Delta_\phi$ on the lines is suppressed). Ellipsis represent $\mathcal{O}(1/N^2)$ graphs which do not contribute to the anomalous dimension $\gamma_\rho$.}
\label{fig:Vrhorho}
\end{figure}

The two-point effective vertex at  order $1/N$ in the presence of the background field $\bar\rho(x)$ is shown in Fig. \ref{fig:Vrhorho1}. It is finite and contributes to $A_\rho$ only; we will calculate it in the next section. Here our aim is to evaluate the anomalous dimension $\gamma_\rho$. We therefore proceed to the $\mathcal{O}(1/N^2)$ diagrams which contribute to the effective two-point vertex. They can be obtained, for example, by dressing the vertices and propagators in Fig. \ref{fig:Vrhorho1}. However, such diagrams are either finite or vanish, because the $\rho$ propagator has no divergences at  order $1/N$, whereas, as we argued in the previous subsection, the next-to-leading order correction to the cubic vertex vanishes at the fixed point. Hence, the only divergent diagrams which contribute to $\gamma_\rho$ are shown in Fig. \ref{fig:Vrhorho}.

Using the Feynman rules  yields,
\begin{eqnarray}
- V_{\rho\rho}^{(1/N^2)} &=&{N\over 4} \Big( {-g_6\over \sqrt{N}} \Big)^2\Big( {2\over \sqrt{N}} \Big)^4 C_\phi^4 C_\rho^2 \left(\int \prod_{i=1}^4 d^3 x_i  ~
  {\bar\sigma(x_1)\bar\sigma(x_2) \over |x_{12}| |x_{13}| |x_{34}|^5 |x_{24}| } \right.
  \nonumber
  \\
 &+& \left. {1\over 2} \int \prod_{i=1}^4 d^3 x_i  ~
  {\bar\sigma(x_1)\bar\sigma(x_2) \over |x_{13}| |x_{23}|  |x_{14}| |x_{24}| |x_{34}|^4} \right) + \ldots~.
  \label{Vrhorho}
\end{eqnarray}
The integral over $x_4$ in the first expression within the parenthesis is power law divergent in the vicinity of $x_4\approx x_3$. In principle, it could have a logarithmic divergence, but its coefficient vanishes; this can be seen by  introducing a spherical sharp cut-off $\mu$, and expanding $|x_4-x_2|^{-1}$ around $x_4=x_3$. 

The power law divergent terms  do not contribute to $\gamma_\rho$, while the finite part of the integral over $x_4$ can be evaluated by analytic continuation in the scaling dimension of the field $\rho$, or by dimensional regularization. In both regularization schemes, one can use the propagator merging relation and remove the regulator at the end. This procedure is sufficient to recover the contribution to $\gamma_\rho$.
The result is, 
\begin{equation}
 \int d^3 x_4 {1\over  |x_{34}|^5 |x_{42}|}= - {2\pi\over 3} {1\over |x_{32}|^3}~.
\end{equation}
Similarly, the integral over $x_4$ in the second expression within the parenthesis in (\ref{Vrhorho}) can be calculated by using the star-triangle relation (\ref{uniqueness scalar}),
\begin{equation}
\int d^3 x_4  ~
  {1\over |x_{14}| |x_{24}||x_{34}|^4} = {-2\pi  |x_{12}| \over |x_{13}|^2|x_{23}|^2}~.
\end{equation}
Hence, 
\begin{equation}
V_{\rho\rho}^{(1/N^2)} ={6\over N^2 \pi^7} \left(\int \prod_{i=1}^3 d^3 x_i  ~
  {\bar\sigma(x_1)\bar\sigma(x_2) \over |x_{12}| |x_{13}| |x_{23}|^3 } + {3\over 2}  \int \prod_{i=1}^3 d^3 x_i  ~
  {\bar\sigma(x_1)\bar\sigma(x_2)  |x_{12}| \over |x_{13}|^3 |x_{23}|^3 } \right) + \ldots~.
\end{equation}
The remaining integrals over $x_3$ are logarithmically divergent. Using a spherical sharp cut-off $\mu$ yields,
\begin{equation}
\label{divergent integral}
 \int d^3 x_3 {1\over |x_{13}|^3|x_{23}|^3}= {8\pi\over |x_{12}|^3} \log(\mu|x_{12}|) + \ldots
\end{equation}
Thus, 
\begin{equation}
V_{\rho\rho}^{(1/N^2)} ={96\over N^2 \pi^6} \int \prod_{i=1}^2 d^3 x_i  ~
  {\bar\sigma(x_1)\bar\sigma(x_2) \over |x_{12}|^2 } ~ \log(\mu|x_{12}|)+ \ldots~.
\end{equation}
Or, equivalently, substituting (\ref{sigma background}) and using the inverse propagator relation (\ref{inverse propagator relation}), 
we obtain
\begin{equation}
V_{\rho\rho}^{(1/N^2)} ={-3072\over N^2 \pi^6} \int \prod_{i=1}^2 d^3 x_i  ~
  {\bar\rho(x_1)\bar\rho(x_2) \over |x_{12}|^4 } ~ \log(\mu|x_{12}|)+ \ldots~,
\end{equation}
Comparing to (\ref{corrected rho propagator}), we get the anomalous dimension,
\begin{equation}
\label{gamma rho}
 \gamma_\rho= {768\over N^2 \pi^4} + \mathcal{O}(1/N^3) ~. 
\end{equation}
This result does not match \cite{Pisarski:1982vz}.

\section{CFT data for the critical $\phi^6$ model}
\label{sec:CFT data for the scalar sextic model}

In this section we use  large $N$ techniques to derive the OPE coefficients and anomalous dimensions of various primary operators in the critical $\phi^6$ model. Our calculations extend known results in the literature \cite{Pisarski:1982vz}.

\subsection{Anomalous dimensions of $(\phi^2)^n$}
\label{sec:next-to-leading order rho propagator}

In the large $N$ limit there are two dynamical fields, $\phi$ and $\rho= \phi^2/\sqrt{N}$,  with scaling dimensions $\Delta_\phi=1/2$ and $\Delta_\rho=1$,  respectively. Their propagators are given by  (\ref{leading order phi propagator}) and (\ref{rho propagator}). In contrast, the dynamics of the field $\sigma\sim \rho^2/\sqrt{N}$ is suppressed by $1/N$.  

As we saw in the previous section,  higher order corrections induce anomalous dimensions to the leading order scalings of the fields. In this subsection, we  calculate the anomalous dimension of the composite operators $(\phi^2)^n$.  The subleading terms in the propagator, both  the anomalous dimension and the amplitude corrections, also contribute to various next-to-leading order three-point functions, and will be needed for the computation of the OPE coefficients, to be performed in the next subsection.  

The full propagator of the field $\rho$ is given by (\ref{corrected rho propagator}). The amplitude correction $A_\rho$ at the next-to-leading order is entirely fixed by attaching two leading-order propagators (\ref{rho propagator}) to the one-loop diagram shown earlier in Fig. \ref{fig:Vrhorho1}. As a result, the $\mathcal{O}(1/N)$ correction to (\ref{rho propagator}) is given by,
\begin{equation}
 \delta \langle\rho(x_3) \rho(0) \rangle= \frac{1}{2}\,\left(\frac{-g_6}{\sqrt{N}}\right)^2\,C_\rho^4 \int {d^3x_1d^3x_2\over |x_1|^{2\Delta_\rho}|x_{12}|^{4\Delta_\rho}|x_{32}|^{2\Delta_\rho}}~.
\end{equation}
The integrals can be evaluated by using (\ref{inverse propagator relation}). At the critical point, $g_6^*=192$, we obtain,
\begin{equation}
A_\rho = \frac{1}{2}\,\left(\frac{-g_6}{\sqrt{N}}\right)^2\,C_\rho^3\,(-2\pi^4) = -\frac{g_6^2}{512\pi^2\,N}=-\frac{72}{\pi^2\,N}\,, \quad  \ \ \ \ \gamma_\rho= 0 + \mathcal{O}(1/N^2)~.
\label{Arho}
\end{equation}

While the anomalous dimension of $\rho$ is highly suppressed, this is not the case for $\rho^n\sim (\phi^2)^n$ with $n\geq 2$.  To leading order in $1/N$ we have
\begin{equation}
\label{rho n propagator}
\langle \rho(x)^n\rho(0)^n\rangle = n!\,\frac{C_\rho^n}{|x|^{2 n  }}\,.
\end{equation}
Or, diagrammatically,
\begin{center}
\scalebox{0.6}{
  \begin{picture}(198,163) (69,-11)
    \SetWidth{1.0}
    \SetColor{Black}
    \Arc[clock](161,-71.875)(161.875,124.205,55.795)
    \Arc[clock](161,46.727)(92.273,170.473,9.527)
    \Vertex(161,104){1}
    \Vertex(161,114){1}
    \Vertex(161,124){1}
    \Text(155,142)[lb]{\scalebox{1.1}{$2\Delta_\rho$}}
    \Text(155,67)[lb]{\scalebox{1.1}{$2\Delta_\rho$}}
    \Text(155,-5)[lb]{\scalebox{1.1}{$2\Delta_\rho$}}
    \Line[](70,62)(252,62)
    \Arc[](161,86.15)(94.15,-165.137,-14.863)
    \CBox(63,55)(77,69){Black}{Black}
    \CBox(245,55)(259,69){Black}{Black}
  \end{picture}
  }
\end{center}
where the black squares stand for  insertions of $\rho^n$, and dots represent additional lines (since $n$ is general). The full propagator, which includes subleading corrections to (\ref{rho n propagator}),  will take the form,
\begin{equation}
\label{next to leading rho n propagator}
\langle \rho(x)^n\rho(0)^n\rangle = n!\, \mu^{-2\gamma_n}
\frac{C_\rho^n (1+A_n)}{|x|^{2(n +\gamma_n)}}\,,
\end{equation}
where $A_n$ and $\gamma_n$ are the $1/N$ corrections to the amplitude and the anomalous dimension, respectively.

\begin{figure}[t!]
\centering \noindent
\includegraphics[width=14cm]{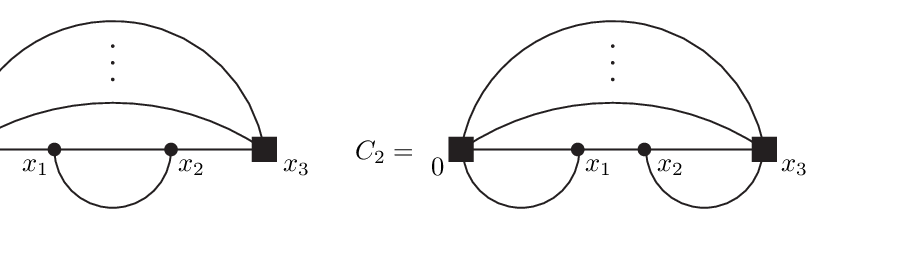}
\newline \caption[]{Finite diagrams which contribute to the full correlator $\langle\rho^n(x_3) \rho^n(0)\rangle$ at  order $1/N$.  The black squares denote insertions of the composite operator $\rho^n$. All propagators are propagators of $\rho$,  (\ref{rho propagator}).}
\label{fig:Cfinite}
\end{figure}

There are four diagrams which contribute to the full propagator at order  $1/N$. Two of them are shown in Fig.~\ref{fig:Cfinite}. They are finite and contribute only to $A_n$,
\begin{eqnarray}
  C_1&=&\frac{n\,n!}{2}\Big({-g_6\over \sqrt{N}}\Big)^2 {C_\rho^{n+3}\over |x_3|^{2 (n-1)}} \int {d^3x_1d^3x_2\over |x_1|^{2 }|x_{12}|^{4 }|x_{32}|^{2 }} ~,
  \nonumber\\
  C_2&=& \frac{n(n-1)\,n!}{2}\Big({-g_6\over \sqrt{N}}\Big)^2 {C_\rho^{n+3}\over |x_3|^{2 (n-2)}} 
  \int {d^3x_1d^3x_2\over |x_1|^{4 }|x_{12}|^{2 }|x_{32}|^{4 }}~.
\end{eqnarray}
The one-loop subdiagrams of $C_1$ and $C_2$ can be evaluated using (\ref{inverse propagator relation}), to get,
\begin{equation}
\label{C1 rhon answer}
C_1 +C_2 = n^2\,n!\frac{C_\rho^n}{|x_3|^{2n }}\, A_\rho~,
\end{equation}
where $A_\rho$ is given by (\ref{Arho}). 

\begin{figure}[t!]
\centering \noindent
\includegraphics[width=12cm]{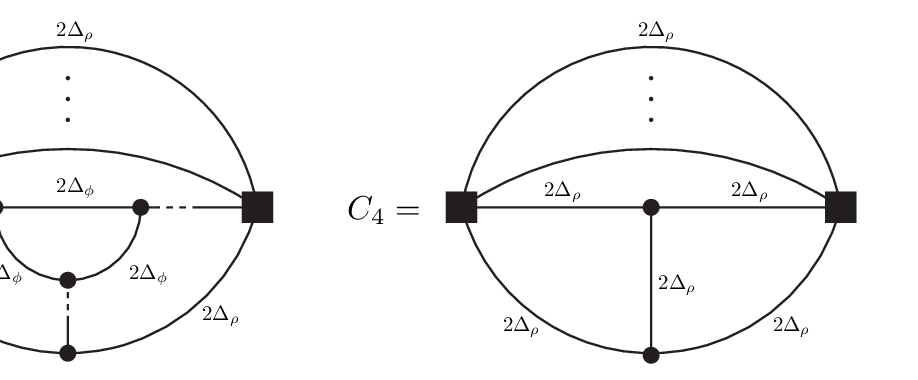}
\newline \caption[]{Divergent diagrams which contribute to the full correlator $\langle\rho^n(x_3) \rho^n(0)\rangle$ at  order $1/N$. The black squares denote insertions of the composite operator $\rho^n$.}
\label{fig:Cdiv}
\end{figure}

In addition, there are two divergent diagrams shown in Fig.~\ref{fig:Cdiv}. These diagrams are responsible for the anomalous dimension. Applying the Feynman rules gives,
\begin{eqnarray}
  C_3&=&-n(n-1)\,n!\,N\Big(\frac{-2}{\sqrt{N}}\Big)^3  \Big({-g_6\over \sqrt{N}}\Big)
   {C_\rho^n C_\phi^3\over |x_3|^{2n-3}}
     \int {d^3x_2\over |x_2|^3 |x_{23}|^3} ~,
  \nonumber\\
  C_4&=& \frac{n(n-1)\,n!}{2}\Big({-g_6\over \sqrt{N}}\Big)^2 {C_\rho^{n+3}\over |x_3|^{2(n-2)}}
  \int {d^3x_1d^3x_2\over |x_1|^2 |x_2|^2 |x_{12}|^2 |x_{13}|^2 |x_{23}|^2}~.
\end{eqnarray}
These expressions can be combined after carrying out the integral over $x_3$, through the use of the star-triangle relation (\ref{uniqueness scalar}),
\begin{equation}
 C_3+C_4= - n(n-1)\,n!\,{8 g_6\over N}\Big(1-{g_6\over 128} \Big)  {C_\rho^n C_\phi^3\over |x_3|^{2n-3}}
 \int {d^3x_2\over |x_2|^3 |x_{23}|^3} ~.
\end{equation}
The integral over $x_2$ diverges logarithmically. This divergence is associated with the anomalous dimension of $\rho^n$. Using (\ref{divergent integral}) yields, 
\begin{equation}
\label{C3 rhon answer}
C_3+C_4 =   - n(n-1)\,n!\,{g_6\over \pi^2 N}C_\rho^n \Big(1-{g_6\over 128} \Big)\,\frac{1}{|x_3|^{2n}}
\log(\mu|x_3|) + \ldots~.
\end{equation}

Combining (\ref{C1 rhon answer}) and (\ref{C3 rhon answer}) with the leading order propagator (\ref{rho n propagator}), and matching the result
with the general form (\ref{next to leading rho n propagator}), gives 
\begin{align}
\gamma_n &= \frac{n(n-1)\,g_6}{ 2 \pi^2 N}\Big(1-{g_6\over 128} \Big)+{\cal O}(1/N^2)
=-\frac{48n(n-1)}{\pi^2 N}+{\cal O}(1/N^2)\,.
\label{gamma rho n}
\end{align}
The anomalous dimensions for $n=1,2$ match \cite{Pisarski:1982vz},\footnote{In fact, $\gamma_2$ in \cite{Pisarski:1982vz} has the opposite sign.} while the anomalous dimensions $\gamma_n$ for $n\geq 3$ are new. As an additional check of our result, we note that the anomalous dimension of $\phi^6$ indeed satisfies $\gamma_3 ={\partial \beta\over \partial g_6}= -{288\over \pi^2 N}$.

\subsection{OPE coefficients}
\label{OPE}

In this section we derive several OPE coefficients, up to the next-to-leading order in $1/N$.

\subsubsection*{Evaluation of $\langle\phi^2 \, \phi^2 \, \phi^2\rangle$ }

\begin{figure}[t!]
\centering \noindent
\hfill\includegraphics[width=12cm]{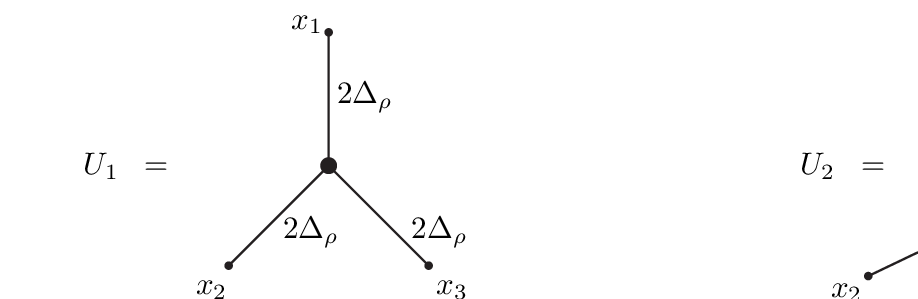}\hspace*{\fill}
\newline \caption[]{Leading order Feynman diagrams of the three-point function $\langle\rho\rho\rho\rangle$. }
\label{fig:U}
\end{figure}

There are two Feynman diagrams which contribute to the leading order three-point function $\langle\rho\rho\rho\rangle$, as shown in Fig.~\ref{fig:U}. Using the star-triangle relation (\ref{uniqueness scalar}) to integrate over the cubic vertex in $U_1$ yields,
\begin{align}
\label{U1 answer}
U_1 = -\frac{g_6 C_\rho^3}{\sqrt{N}}\,\int d^3x_4\,\frac{1}{(|x_{14}||x_{24}||x_{34}|)^2}
= -\frac{g_6}{(8\pi)^3\,\sqrt{N}}\,\frac{1}{|x_{12}||x_{13}||x_{23}|}~.
\end{align}
The other diagram, $U_2$, is trivial to compute, as all three integrals are taken over delta functions (\ref{rho and sigma correlator}), giving,
\begin{equation}
\label{U2 answer}
U_2 = \frac{8C_\phi^3}{\sqrt{N}}\,\frac{1}{|x_{12}||x_{13}||x_{23}|}
= \frac{1}{8\pi^3\sqrt{N}}\,\frac{1}{|x_{12}||x_{13}||x_{23}|}~.
\end{equation}
Thus, at the fixed point $g_6^*=192$, we get,
\begin{equation}
\label{Non-normalized Crho3}
\langle \rho(x_1)\rho(x_2)\rho(x_3)\rangle =U_1+U_2=\frac{C_{\rho\rho\rho}}{|x_{12}||x_{13}||x_{23}|}~, \quad
C_{\rho\rho\rho}=-\frac{1}{4\pi^3 \sqrt{N}}\Big(1 + \mathcal{O}(1/N)\Big)~.
\end{equation}

It is standard to define the OPE coefficients to be for the fields which are normalized such that the amplitude of the two-point correlation function is unity. Based on (\ref{corrected rho propagator}), we therefore rescale the field $\rho$, 
\begin{equation}
\label{rho normalization}
\rho\rightarrow \sqrt{C_\rho(1+A_\rho)}\,\rho ~,
\end{equation}
and get,
\begin{equation}
\label{leading normalized rho rho rho}
\langle \rho(x_1)\rho(x_2)\rho(x_3)\rangle \Big|_{\textrm{normalized}} = \frac{\hat C_{\rho\rho\rho}}{|x_{12}||x_{13}||x_{23}|}~,
\end{equation}
where the OPE coefficient is given by
\begin{equation}
\label{hatC3 result}
\hat C_{\rho\rho\rho} =- 4{\sqrt{2 \over N}}\Big(1 + \mathcal{O}(1/N)\Big)~.
\end{equation}

The next-to-leading order correction  to this OPE coefficient follows directly from our previous results. In particular, it is entirely determined by the $1/N$ corrections to the external propagators and to the cubic vertex. The former is encoded in the coefficient $A_\rho$ given by (\ref{Arho}), whereas the latter, as argued in section \ref{sec: UV fixed point}, vanishes. Hence, for $\rho$ normalized according to (\ref{rho normalization}), we find that OPE coefficient up to order $1/N$ is,
\begin{equation}
\label{next to-leading normalized rho rho rho}
\hat C_{\rho\rho\rho} =- 4{\sqrt{2 \over N}}\Big(1 + {3\over 2} A_\rho + \mathcal{O}(1/N^2)\Big)~.
\end{equation}

\subsubsection*{Evaluation of $\langle\phi \, \phi \, \phi^2\rangle$}

We now consider the three-point function $\langle\phi \, \phi \, \phi^2\rangle \sim \langle\phi \phi \rho\rangle$. To leading order in $1/N$ it is determined by a tree diagram of the form 
\begin{center}
  \begin{picture}(180,120) (39,-20)
    \SetWidth{1.0}
    \SetColor{Black}
    \Vertex(40,42){2}
    \Line[](40,42)(72,42)
    \Line[dash,dashsize=5](72,42)(112,42)
    \Line[](112,42)(150,80)
    \Line[](112,42)(150,-4)
    \Vertex(112,42){4}
    \Vertex(150,80){2}
    \Vertex(150,-4){2}
    \Text(40,50)[lb]{$x_3$}
    \Text(155,80)[lb]{$x_1$}
    \Text(155,-4)[lb]{$x_2$}
    \Text(135,53)[lb]{\scalebox{0.8}{$2\Delta_\phi$}}
    \Text(135,16)[lb]{\scalebox{0.8}{$2\Delta_\phi$}}
    \Text(182,35)[lb]{$=\quad\frac{2}{\sqrt{N}}\,C_\phi^2\, {1\over |x_{13}| |x_{23}|} ~,$}
  \end{picture}
\end{center}
where we  used the Feynman rules of section \ref{sec:preliminaries} to get the expression on the right hand side.

\begin{figure}[t!]
\centering \noindent
\hfill\includegraphics[width=4cm]{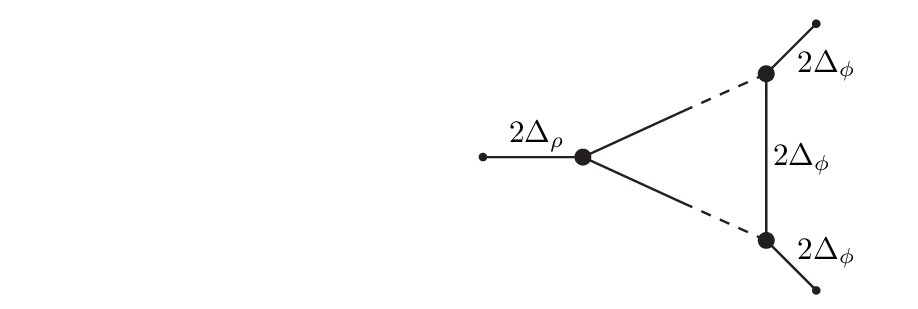}\hspace*{\fill}
\newline \caption[]{The $\mathcal{O}(1/N)$ correction to the three-point function $\langle\phi \, \phi \, \rho\rangle$. }
\label{fig:phiphirho}
\end{figure}

The next-to-leading order correction is obtained by accounting for the $1/N$ corrections to the external propagators and the $\rho \phi^2$ vertex. The correction to the cubic vertex is represented by the one-loop diagram in Fig.~\ref{fig:phiphirho}.
It vanishes, because the integrals over the delta functions (\ref{rho and sigma correlator}) result in $\langle \phi^2\rangle = 0$. 

Furthermore, as argued in section \ref{sec:next-to-leading order rho propagator}, only the amplitude of the $\rho$ propagator is modified at the next-to-leading order, whereas the anomalous dimension and the $1/N$ correction to the $\phi$ propagator are absent at this order. As a result, the three-point function $\langle\phi \phi \rho\rangle$ up to order $1/N$  is given by,
\begin{equation}
 \langle\phi(x_1) \phi(x_2) \rho(x_3)\rangle = \frac{2}{\sqrt{N}}\, {C_\phi^2 \over |x_{13}| |x_{23}|}\Big(1 + A_\rho +  \mathcal{O}(1/N^2)\Big)~.
\end{equation}
To get the OPE coefficient, we want to first properly normalize the fields: we rescale $\rho$, as was done in (\ref{rho normalization}), and similarly rescale $\phi\to\sqrt{C_\phi(1+A_\phi)} \phi$, see (\ref{corrected phi propagator}). Since $A_\phi \sim \mathcal{O}(1/N^2)$ and $A_\rho$ is given by (\ref{Arho}), we get
\begin{equation}
\label{phi phi rho general}
\langle \phi(x_1)\phi(x_2)\rho(x_3)\rangle \Big|_{\textrm{normalized}} 
= \frac{\hat C_{\phi\phi\rho}}{|x_{13}||x_{23}|}~, \quad \ \ \  \hat C_{\phi\phi\rho}= \sqrt{2\over N} \Big(1 -\frac{36}{\pi^2\,N}  +\mathcal{O}(1/N^2)\Big)~.
\end{equation}
Notice that (\ref{phi phi rho general}) is conformally covariant, because $2\Delta_\phi - \Delta_\rho = 0$ and, as argued in section~\ref{known anomalous dims}, $\gamma_{\phi} \sim \gamma_\rho \sim {\cal O}(1/N^2)$. 

\section{Review of the Gross-Neveu model}
\label{sec:preliminaries gn}

In this section we review some aspects of the Gross-Neveu model \cite{Gross:1974jv} which will be needed for section  \ref{sec: cft data in gross-neveu}. Those who are familiar with the subject may proceed directly to the next section. 

Consider the $U(n)$-invariant Gross-Neveu model in $d$-dimensional Euclidean space, 
\begin{align}
\label{starting action}
S = \int d^dx \, \left(  \bar\psi \gamma^\mu \partial _\mu \psi +
\frac{g}{N}\,\left(\bar\psi\psi)^2 \right)
\right)\,,
\end{align}
where $\psi$ collectively denotes $n$ Dirac fermions, and
$\bar\psi   = \psi ^ \dagger$ denotes the standard Hermitian conjugation. We choose conventions with Hermitian $\gamma$-matrices, 
$(\gamma^\mu)^\dagger = \gamma^\mu$, and, as is standard  in large-$N$ fermionic models, take $N = n\,\textrm{tr}\,\mathbb{I}$ where $\mathbb{I}$ is the identity matrix in the space of $2^{[d/2]}$-dimensional Dirac spinors \cite{ZinnJustin:1991yn}.

The standard trick for solving the Gross-Neveu model is to rewrite the model (\ref{starting action})
using the Hubbard-Stratonovich field $s$, 
\begin{align}
\label{starting action HS}
S = \int d^dx \, \left( \bar\psi \gamma^\mu \partial _\mu \psi 
-\frac{1}{4g}\,s^2 + \frac{1}{\sqrt{N}}\,s\bar\psi\psi \right)\,.
\end{align}
The coupling $g$ is irrelevant at the Gaussian fixed point, and the model is therefore IR free in dimensions $d>2$. However, the $\epsilon$-expansion suggests that it has a non-trivial UV fixed point, and the corresponding CFT has been extensively studied \cite{ZinnJustin:1991yn,Gracey:1990wi,Gracey:1992cp,Derkachov:1993uw,Vasiliev:1992wr,Vasiliev:1993pi,Gracey:1993kb,Gracey:1993kc,Manashov:2016uam,Manashov:2017rrx}. In strictly two dimensions, the model (\ref{starting action HS}) is a renowned example of an asymptotically free field theory \cite{Gross:1974jv}.\footnote{See \cite{Gracey:2016mio} and references therein for four-loop renormalization of the two-dimensional Gross-Neveu model.} As we will review in the next subsection, the large-$N$ scaling dimension of $s$ at the UV fixed point is equal to one, and so the corresponding CFT is unitary provided that $d\leq 4$.

\subsection{Feynman rules and useful identities}
\label{sec:useful identities}

In momentum space the bare propagators of the fields in (\ref{starting action HS}) are given by,~\footnote{We will generally  not keep track of the $U(n)$ indices.}
\begin{align}
\label{momentum space propagators}
\langle s(p)s(q)\rangle_\text{bare} = (2\pi)^d\,\delta ^{(d)}(p+q)\,(-2g)\,,\quad
\langle \psi(p)\bar\psi(q)\rangle = (2\pi)^d\,\delta^{(d)}(p+q)\,\frac{-ip^\mu\gamma^\mu}{p^2}\,,
\end{align}
Since we are interested in the large-$N$ limit of the model, we have to resum the so-called bubble diagrams to get the leading order propagator of the auxiliary field $s$ \cite{Gross:1974jv},
\begin{equation}
\label{Gs general}
\langle s(p)s(q)\rangle_{N\to\infty}  = -2g\,\sum_{n=0}^\infty (-2gB(p))^n = \frac{-2g}{1+2gB(p)}\quad \underset{\text{UV}}{\to} \quad -\frac{1}{B(p)} \,.
\end{equation}
In the last step we took the limit of large momentum, to ensure that the model is sitting at the UV fixed point. The fermionic bubble $B(p)$ appearing above is given by,
\begin{equation}
B(p)=\frac{n}{N}\,\textrm{tr}\,\int \frac{d^dq}{(2\pi)^d}\,
\frac{-i\gamma^\mu q_\mu (-i)\gamma^\nu (p-q)_\nu}{q^2(p-q)^2}=\frac{p^{d-2}}{4^{d-1}\pi^\frac{d-3}{2}\,\sin\left(\frac{\pi d}{2}\right)
\Gamma\left(\frac{d-1}{2}\right)}\,.
\end{equation}
Using the standard Fourier transform relation,
\begin{equation}
\label{A of Delta def}
\int \frac{d^dx}{(2\pi)^d}\,e^{ik\cdot x}\,\frac{1}{|k|^{d-2\Delta}}
=\frac{2^{2\Delta-d}}{\pi^{d/2}}\,\frac{1}{A(\Delta)}\,\frac{1}{|x|^{2\Delta}} ~,
\quad \ \ \ \ \  A(\Delta) = \frac{\Gamma\left(\frac{d}{2}-\Delta\right)}{\Gamma(\Delta)}~,
\end{equation}
we get the conformal propagator of the auxiliary field $s$ in position space \cite{ZinnJustin:1991yn} (see also \cite{Manashov:2016uam,Manashov:2017rrx}),
\begin{equation}
\label{Cs}
\langle s(x)s(0)\rangle = \frac{C_s}{|x|^{2\Delta_s}}~, \quad C_s = -\frac{2^d\sin\left(\frac{\pi d}{2}\right)
\Gamma\left(\frac{d-1}{2}\right)}{\pi^{\frac{3}{2}}\Gamma\left(\frac{d}{2}-1\right)} ~, \quad \Delta_s=1~.
\end{equation}
Similarly, the propagator of the fermion field $\psi$ is,
\begin{equation}
\label{psiprop}
\langle \psi(x)\bar\psi (0)\rangle =C_\psi\,\frac{x^\mu\gamma_\mu}{|x|^{2\Delta_\psi+1}}~, \quad C_\psi = \frac{\Gamma\left(\frac{d}{2}\right)}{2\pi^\frac{d}{2}}~, 
\quad \Delta _\psi = \frac{d-1}{2}\,.
\end{equation}

\begin{figure}[t!]
\centering \noindent
\hfill\includegraphics[width=12cm]{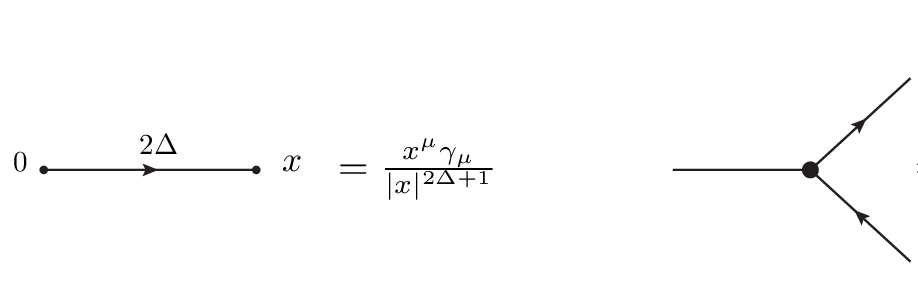}\hspace*{\fill}
\newline \caption[]{Feynman rules in the presence of Dirac field.}
\label{fig:Dirac_rules}
\end{figure}

We will use position space Feynman rules, as we did in the previous sections for the scalar field theory. Since we now have fermions, a Dirac propagator will have an arrow, and there is no symmetry factor of $2$ in the $\bar\psi \psi s$ vertex, see Fig. \ref{fig:Dirac_rules}. The amplitudes of the propagators are normalized to unity, and therefore one needs to multiply each Feynman diagram by an appropriate power of $C_{\psi,s}$. 

The integral identities that were used in scalar conformal perturbation theory can be generalized to accommodate fermions 
\cite{Gracey:1990wi} (see also \cite{Preti:2018vog} for a recent review), {\it e.g.,}
\begin{center}
\scalebox{0.9}{
  \begin{picture}(197,71) (39,-5)
    \SetWidth{1.0}
    \SetColor{Black}
    \Vertex(40,31){2}
    \Arc[arrow,arrowpos=0.5,arrowlength=5,arrowwidth=2,arrowinset=0.2,clock](82.5,-4)(55.057,141,39)
    \Vertex(125,31){2}
    \Arc[clock](82.5,65)(55.057,-39,-141)
    \Text(135,28)[lb]{$=$}
    \Vertex(155,31){2}
    \Line[arrow,arrowpos=0.5,arrowlength=5,arrowwidth=2,arrowinset=0.2](155,31)(215,31)
    \Vertex(215,31){2}
    \Text(165,37)[lb]{\scalebox{0.8}{$2\Delta_1+2\Delta_2$}}
    \Text(80,57)[lb]{\scalebox{0.8}{$2\Delta_1$}}
    \Text(80,-3)[lb]{\scalebox{0.8}{$2\Delta_2$}}
  \end{picture}
  }
\end{center}
\begin{center}
\scalebox{0.9}{
  \begin{picture}(197,71) (39,-5)
    \SetWidth{1.0}
    \SetColor{Black}
    \Vertex(40,31){2}
    \Arc[arrow,arrowpos=0.5,arrowlength=5,arrowwidth=2,arrowinset=0.2,clock](82.5,-4)(55.057,141,39)
    \Vertex(125,31){2}
    \Arc[arrow,arrowpos=0.5,arrowlength=5,arrowwidth=2,arrowinset=0.2,clock](82.5,65)(55.057,-39,-141)
    \Text(135,28)[lb]{$=$}
    \Vertex(155,31){2}
    \Line[](155,31)(215,31)
    \Vertex(215,31){2}
    \Text(165,37)[lb]{\scalebox{0.8}{$2\Delta_1+2\Delta_2$}}
    \Text(80,57)[lb]{\scalebox{0.8}{$2\Delta_1$}}
    \Text(80,-3)[lb]{\scalebox{0.8}{$2\Delta_2$}}
    \Text(225,27)[lb]{$\times\;\; (-\mathbb{I})$}
  \end{picture}
  }
\end{center}
Moreover, the propagator merging relations with fermions take the form,
\begin{center}
\scalebox{0.9}{
  \begin{picture}(271,37) (49,-10)
    \SetWidth{1.0}
    \SetColor{Black}
    \Vertex(-20,7){2}
    \Line[arrow,arrowpos=0.5,arrowlength=5,arrowwidth=2,arrowinset=0.2](-20,7)(35,7)
    \Line[](35,7)(85,7)
    \Vertex(85,7){2}
    \Vertex(35,7){4}
    \Text(2,12)[lb]{\scalebox{0.8}{$2\Delta_1$}}
    \Text(57,12)[lb]{\scalebox{0.8}{$2\Delta_2$}}
    \Text(101,4)[lb]{$=$}
    \Vertex(125,7){2}
    \Line[arrow,arrowpos=0.5,arrowlength=5,arrowwidth=2,arrowinset=0.2](125,7)(210,7)
    \Vertex(210,7){2}
    \Text(140,12)[lb]{\scalebox{0.8}{$2(\Delta_1+\Delta_2)-d$}}
    \Text(215,2)[lb]{\scalebox{1}{$~\times ~ \pi^\frac{d}{2}\,A(\Delta_2)V(\Delta_1,d-\Delta_1-\Delta_2)$}}
  \end{picture}
  }
\end{center}
\begin{center}
\scalebox{0.9}{
  \begin{picture}(271,37) (49,-10)
    \SetWidth{1.0}
    \SetColor{Black}
    \Vertex(-20,7){2}
    \Line[arrow,arrowpos=0.5,arrowlength=5,arrowwidth=2,arrowinset=0.2](-20,7)(35,7)
    \Line[arrow,arrowpos=0.5,arrowlength=5,arrowwidth=2,arrowinset=0.2](35,7)(85,7)
    \Vertex(85,7){2}
    \Vertex(35,7){4}
    \Text(2,12)[lb]{\scalebox{0.8}{$2\Delta_1$}}
    \Text(57,12)[lb]{\scalebox{0.8}{$2\Delta_2$}}
    \Text(101,4)[lb]{$=$}
    \Vertex(125,7){2}
    \Line[](125,7)(210,7)
    \Vertex(210,7){2}
    \Text(140,12)[lb]{\scalebox{0.8}{$2(\Delta_1+\Delta_2)-d$}}
    \Text(215,4)[lb]{\scalebox{1}{$~\times~ (-\pi^\frac{d}{2})\,A(d-\Delta_1-\Delta_2) V(\Delta_1,\Delta_2)\times \mathbb{I}$}}
  \end{picture}
  }
\end{center}
where the insertion point of the middle vertex on the left-hand side is integrated over, $A(\Delta)$
is defined in (\ref{A of Delta def}), and 
\begin{align}
V(\Delta_1,\Delta_2) &= \frac{\Gamma\left(\frac{d}{2}-\Delta_1+\frac{1}{2}\right)}{\Gamma(\Delta_1+\frac{1}{2})}
\frac{\Gamma\left(\frac{d}{2}-\Delta_2+\frac{1}{2}\right)}{\Gamma(\Delta_2+\frac{1}{2})}\,.
\end{align}
In the case of a Yukawa vertex, the star-triangle relation (\ref{uniqueness scalar}) becomes ($\Delta_1+\Delta_2+\Delta_3 = d$),
\begin{equation}
 \label{star-triangle}
  \int d^dx_4\,\frac{\gamma_\mu x_{41}^\mu \gamma_\nu x_{24}^\nu}
  {|x_{14}|^{2\Delta_1+1}|x_{24}|^{2\Delta_2+1}|x_{34}|^{2\Delta_3}}= \frac{\pi^\frac{d}{2}A(\Delta_3)V(\Delta_1,\Delta_2) ~\gamma_\mu x_{31}^\mu
\gamma_\nu x_{23}^\nu}{|x_{12}|^{d-2\Delta_3}|x_{13}|^{d-2\Delta_2+1}|x_{23}|^{d-2\Delta_1+1}} ~,
\end{equation}
or diagrammatically, 
\begin{center}
\scalebox{0.8}{
  \begin{picture}(390,106) (29,-27)
    \SetWidth{1.0}
    \SetColor{Black}
    \Line[arrow,arrowpos=0.5,arrowlength=5,arrowwidth=2,arrowinset=0.2](96,28)(144,76)
    \Line[arrow,arrowpos=0.5,arrowlength=5,arrowwidth=2,arrowinset=0.2](144,-26)(96,28)
    \Vertex(144,-26){2}
    \Vertex(30,28){2}
    \Vertex(144,76){2}
    \Line[](96,28)(30,28)
    \Vertex(96,28){4}
    \Line[arrow,arrowpos=0.5,arrowlength=5,arrowwidth=2,arrowinset=0.2](196,28)(310,76)
    \Line[arrow,arrowpos=0.5,arrowlength=5,arrowwidth=2,arrowinset=0.2](310,-26)(196,28)
    \Line[](310,76)(310,-26)
    \Text(166,26)[lb]{\scalebox{1}{$=$}}
    \Text(324,22)[lb]{\scalebox{1}{$\times ~ \pi^\frac{d}{2}A(\Delta_3)V(\Delta_1,\Delta_2)$}}
    \Text(60,32)[lb]{\scalebox{0.8}{$2\Delta_3$}}
    \Text(103,58)[lb]{\scalebox{0.8}{$2\Delta_2$}}
    \Text(103,-8)[lb]{\scalebox{0.8}{$2\Delta_1$}}
    \Text(233,60)[lb]{\scalebox{0.8}{$d-2\Delta_1$}}
    \Text(233,-18)[lb]{\scalebox{0.8}{$d-2\Delta_2$}}
    \Text(275,22)[lb]{\scalebox{0.8}{$d-2\Delta_3$}}
  \end{picture}
  }
\end{center}

\subsection{Propagators at the next-to-leading order}
\label{sec:corrections to propagators}

We now review the formalism to reproduce the leading order anomalous dimensions of the fields $s$ and $\psi$, as well as the $1/N$ corrections to the amplitudes of their propagators \cite{Gracey:1990wi,Manashov:2017rrx}. These results will be  used extensively in the next section.   

For the Dirac fermion the relevant conformal graph is,
\begin{center}
\scalebox{0.8}{
  \begin{picture}(296,100) (-20,-2)
    \SetWidth{1.0}
    \SetColor{Black}
    \Text(-30,39)[lb]{\scalebox{1.2}{$
    \mu^{-\delta}\;\;\times$}}
    \Arc[arrow,arrowpos=0.5,arrowlength=5,arrowwidth=2,arrowinset=0.2](129.536,43.535)(40.539,179.343,363.486)
    \Arc[](128.994,43.994)(41.006,0.008,181.389)
    \Line[arrow,arrowpos=0.5,arrowlength=5,arrowwidth=2,arrowinset=0.2](25,44)(88,44)
    \Line[arrow,arrowpos=0.5,arrowlength=5,arrowwidth=2,arrowinset=0.2](170,43)(234,44)
    \Vertex(87,44){4}
    \Vertex(170,43){4}
    \Vertex(25,44){2}
    \Vertex(232,44){2}
    \Text(18,50)[lb]{\scalebox{0.8}{$x_2$}}
    \Text(231,50)[lb]{\scalebox{0.8}{$x_1$}}
    \Text(75,51)[lb]{\scalebox{0.8}{$x_4$}}
    \Text(175,51)[lb]{\scalebox{0.8}{$x_3$}}
    \Text(45,52)[lb]{\scalebox{0.8}{$d-1$}}
    \Text(195,51)[lb]{\scalebox{0.8}{$d-1$}}
    \Text(120,92)[lb]{\scalebox{0.8}{$2+\delta$}}
    \Text(120,-10)[lb]{\scalebox{0.8}{$d-1$}}
    \Text(250,33)[lb]{\scalebox{1.2}{$=C_\psi \, \hat P\,\frac{\gamma_\mu x_{12}^\mu}{|x_{12}|^{d}}
    \frac{1}{(|x_{12}|\mu)^\delta}$}}
  \end{picture}
}
\end{center}
This diagram diverges; we regularized it by introducing a slight shift $\delta$ in the scaling dimension of $s$ and an arbitrary scale $\mu$, to maintain correct dimensionality \cite{Parisi:1972zm,Vasiliev:1975mq}.
The integrals over internal vertices $x_{3,4}$ can be done by using the propagator
merging relations reviewed above. Expanding around $\delta = 0$ yields,
\begin{equation}
\hat P = \frac{2\gamma_\psi}{\delta} + A_\psi\,,
\end{equation}
where \cite{Gracey:1990wi,Manashov:2017rrx},
\begin{align}
\gamma_\psi &=-\frac{1}{N}\frac{2^{d-1} \sin \left(\frac{\pi  d}{2}\right)
\Gamma \left(\frac{d-1}{2}\right)}{\pi ^{3/2} d \Gamma \left(\frac{d}{2}-1\right)}\,,\\
A_\psi &= -\frac{2}{d}\,\gamma_\psi\,.
\end{align}
Similarly, one can derive the regularized $s$ field propagator \cite{Manashov:2017rrx}
\begin{equation}
\langle s(x) s(0)\rangle =C_s\frac{1}{|x|^2}\frac{1}{(|x|\mu)^\delta}\left(1
+\frac{2\gamma_s}{\delta} +A_s\right)\,,
\end{equation}
with 
\begin{align}
\gamma_s &=\frac{1}{N}\frac{4 \sin \left(\frac{\pi  d}{2}\right) \Gamma (d)}{\pi  d \Gamma \left(\frac{d}{2}\right)^2}
=-4\,\frac{d-1}{d-2}\,\gamma_\psi\\
A_s&=-\left(H_{d-2}+\frac{2}{d}+\pi  \cot \left(\frac{\pi  d}{2}\right)\right)\,\gamma_s\,,
\end{align}
where $H_n$ is the $n$'th harmonic number.

To remove the $1/\delta$ poles in the above propagators, one has to renormalize the fields,
\begin{equation}
\label{fields renormalization}
\psi\rightarrow \sqrt{1+\frac{2\gamma_\psi}{\delta}}\,\psi\,,\qquad
s\rightarrow \sqrt{1+\frac{2\gamma_s}{\delta}}\,s\,.
\end{equation}
As a result, the physical correlators take the form,~\footnote{In order to avoid cluttering the notation, we do not distinguish between physical and bare fields.}
\begin{align}
\label{full psi propagator}
\langle \psi(x)\bar\psi(0)\rangle  &= C_\psi \, 
(1+A_\psi)\,\mu^{-2\gamma_\psi}\,\frac{x^\mu\gamma_\mu}{|x|^{2(\Delta_\psi+\gamma_\psi)+1}}\,,\\
\label{full s propagator}
\langle s(x)s(0)\rangle &= C_s\,(1+A_s)\,\mu^{-2\gamma_s}\,\frac{1}{|x|^{2(\Delta_s+\gamma_s)}}\,.
\end{align}

Notice that the field strength renormalization (\ref{fields renormalization}) induces a 
counterterm, which at the next-to-leading order in $1/N$ is given by,
\begin{equation}
\frac{1}{\sqrt{N}}\,\bar\psi\psi s\rightarrow \frac{1}{\sqrt{N}}\,\bar\psi\psi s
+\frac{2\gamma_\psi+\gamma_s}{\delta}\,\frac{1}{\sqrt{N}}\,\bar\psi\psi s~.
\end{equation}
Or, equivalently,
\begin{equation}
\label{vertex counterterm}
S_{\textrm{int}}^{\textrm{c.t.}} = \frac{2\gamma_\psi+\gamma_s}{\delta}\,\frac{1}{\sqrt{N}}\,\int d^dx ~ \bar\psi\psi s\,.
\end{equation}

\subsection{Gross-Neveu-Yukawa model} \label{GNYsub}

Before closing this section, we briefly review the relation between the Gross-Neveu and  Gross-Neveu-Yukawa (GNY) models. The results reviewed here are used to check our calculations in the next section.

The GNY model is defined by the action,
\begin{equation}
\label{GNY definition}
S= \int d^dx\,\left(\bar\psi \gamma^\mu\partial_\mu\psi +g_1\bar\psi\psi s
+\frac{1}{2}\,(\partial s)^2+\frac{g_2}{24}\,s^4\right)\,.
\end{equation}
This model is manifestly renormalizable in $d\leq4$, because the couplings $g_1$ and $g_2$ are relevant at the Gaussian fixed point in $d<4$ and marginally irrelevant in $d=4$. Below four dimensions they flow to an interacting fixed point in the IR limit. The critical values of the couplings can be calculated by using the standard $\epsilon$-expansion in $d=4-\epsilon$ dimensions \cite{ZinnJustin:1991yn},
\begin{align}
\label{g1star}
g_1^\star &= 4\pi\,\sqrt{\frac{\epsilon}{N+6}} + {\cal O}(\epsilon)
=4\pi\,\sqrt{\frac{\epsilon}{N}}\,\left(1-\frac{3}{N} + {\cal O}\big(1/N^2\big)\right) + {\cal O}(\epsilon)\,,\\
g_2^\star &= \frac{384\pi^2\epsilon \,N}{(N+6)(N-6+\sqrt{N^2+132N+36})}\,.
\end{align}
Furthermore, the scaling dimensions of the fields at the fixed point were calculated in \cite{ZinnJustin:1991yn}
(see \cite{Moshe:2003xn} for a comprehensive review),
\begin{equation}
\Delta_\psi^{(4-\epsilon)} = \frac{3}{2}-\frac{N+4}{2(N+6)}\,\epsilon+{\cal O}(\epsilon^2)\,\qquad
\Delta_s^{(4-\epsilon)} = 1 - \frac{3}{N+6}\,\epsilon+{\cal O}(\epsilon^2)\,.
\end{equation}
This formula shows that in the large $N$ limit $\Delta_s^{(4-\epsilon)} = 1 + {\cal O}(1/N)$, up to linear order in $\epsilon$. In fact, a general argument based on the $1/N$ expansion shows that this relation is true for any $2\leq d \leq 4$, see, \textit{e.g.}, \cite{Moshe:2003xn}. In particular, the operators $(\partial s)^2$ and $s^4$ are irrelevant in the large $N$ limit below four dimensions, because their scaling dimension approaches 4 in this limit. Therefore, the effective action of the GNY model  in the deep IR matches that of the GN model in the deep UV, and the two models are described by the same CFT at their respective fixed points \cite{ZinnJustin:1991yn, Fei:2014yja,Diab:2016spb,Fei:2016sgs}.

As was pointed out in \cite{ZinnJustin:1991yn}, the equivalence between the
GN and the GNY models is analogous to the equivalence between the critical $\phi^4$ vector
model and the non-linear sigma model. A similar equivalence, albeit one that is slightly less exhaustive, also holds between the critical $\phi^4$  vector model in $4<d\leq 6$ and the model of $N + 1$ massless scalars with cubic interaction  \cite{Fei:2014yja,Fei:2014xta,Gracey:2015tta}.\footnote{It was argued  in these works that the equivalence between the models in $d=6-\epsilon$ dimensions holds up to ${\cal O}(\epsilon^4)$ order.}

As an illustration of the equivalence between the critical GN and GNY models, let us consider the 1PI vertices  for the field $s$. To leading order in the $1/N$ expansion, they are represented diagrammatically 
by a single fermionic loop with a given number of $s$ legs attached to it. Because of the equivalence, vertices with the same number and type of legs in both models should match,\footnote{A similar correspondence holds between the critical $\phi^4$ vector model and the model of $N + 1$ massless scalars with cubic interaction \cite{Goykhman:2019kcj}.} {\it i.e.,}  for $n$ external fields $s$, we have
\begin{equation}
 \Big(-\frac{1}{\sqrt{N}}\Big)^n\,C_s^\frac{n}{2}C_\psi^n \,
\int\prod_{i=1}^n d^d x_i \, s_i \, {\cal I}_n(x_1,\dots,x_n)=\left(-g_1^\star\right)^n\,\tilde C_s^\frac{n}{2}C_\psi^n \,
\int\prod_{i=1}^n d^d x_i \, s_i \, {\cal I}_n(x_1,\dots,x_n)\,,
\end{equation}
where the two sides of this equation represent the same vertex in the GN and GNY models respectively, ${\cal I}_n(x_1,\dots,x_n)$ corresponds to the internal $\psi$-loop, and the external fields $s$ are normalized such that the amplitude of their propagators equals unity. In particular, $\tilde C_s = \Gamma\left(\frac{d}{2}-1\right)/(4\pi^\frac{d}{2})$, because the field $s$ in the GNY action (\ref{GNY definition}) is canonically normalized. Using (\ref{Cs}) and (\ref{g1star}), one can verify that the above identity indeed holds to  leading order in the $\epsilon$ and $1/N$ expansion.

Moreover, the symmetry pattern of the models match. The actions (\ref{starting action HS}) and  (\ref{GNY definition}) are invariant under the discrete symmetry \cite{Gross:1974jv,ZinnJustin:1991yn,Moshe:2003xn}
\begin{align}
\label{parity like symmetry}
&(x^1,\dots, x^{\mu-1}, x^\mu, x^{\mu+1},\dots, x^d )\rightarrow
(x^1,\dots, x^{\mu-1}, -x^\mu, x^{\mu+1},\dots, x^d )\,, \\ 
&s\rightarrow -s\,,\quad \psi\rightarrow \gamma_\mu \psi\,,\quad
\bar\psi \rightarrow -\bar\psi \gamma_\mu\,.\notag
\end{align}
Notice that such a transformation leaves the fermionic kinetic term invariant while it transforms $\bar\psi\psi\rightarrow -\bar\psi\psi$,
in analogy with the chiral transformation in even dimensions and parity in odd dimensions.
Thus, for instance, any correlator with an odd number of fields $s$ must vanish. The three point function $\langle sss \rangle$ is the simplest instance where one can directly see that the correlator vanishes. To leading order in the large-$N$ expansion, it is given by,
\begin{center} 
  \begin{picture}(270,137) (23,3)
  \scalebox{0.8}{
    \SetWidth{1.0}
    \SetColor{Black}
    \Line[](181,155)(181,96)
    \Line[arrow,arrowpos=0.5,arrowlength=5,arrowwidth=2,arrowinset=0.2](180,95)(152,47)
    \Line[arrow,arrowpos=0.5,arrowlength=5,arrowwidth=2,arrowinset=0.2](152,47)(208,47)
    \Line[arrow,arrowpos=0.5,arrowlength=5,arrowwidth=2,arrowinset=0.2](211,44)(183,96)
    \Line[](211,47)(260,16)
    \Line[](150,47)(100,16)
    \Vertex(181,97){4}
    \Vertex(152,47){4}
    \Vertex(211,47){4}
    \Vertex(181,156){2}
    \Vertex(100,16){2}
    \Vertex(260,16){2}
    \Text(165,159)[lb]{\scalebox{1.1}{$x_1$}}
    \Text(88,6)[lb]{\scalebox{1.1}{$x_2$}}
    \Text(265,6)[lb]{\scalebox{1.1}{$x_3$}}
    \Text(186,126)[lb]{\scalebox{1.1}{$2$}}
    \Text(120,36)[lb]{\scalebox{1.1}{$2$}}
    \Text(237,36)[lb]{\scalebox{1.1}{$2$}}
    \Text(205,73)[lb]{\scalebox{1.1}{$d-1$}}
    \Text(170,32)[lb]{\scalebox{1.1}{$d-1$}}
    \Text(130,73)[lb]{\scalebox{1.1}{$d-1$}}
    \Text(0,73)[lb]{\scalebox{1.1}{$\langle s(x_1)s(x_2)s(x_3)\rangle =$}}
    \Text(270,73)[lb]{\scalebox{1.1}{$\propto \quad x_{12}^\mu x_{23}^\nu x_{31}^\lambda\,
    \textrm{tr}(\gamma_\mu\gamma_\nu\gamma_\lambda)=0 \, ,$}}
    }
      \end{picture}
\end{center}
where the last equality is true in any integer dimension $d=2,3,4$, whereas in general $2<d<4$ it holds by analytic continuation \cite{Manashov:2016uam}.\footnote{The trace of the gamma matrices vanishes identically in even dimensions. In $d=3$, $\textrm{tr}(\gamma_\mu\gamma_\nu\gamma_\lambda)\propto \epsilon^{\mu\nu\lambda}$.}

In contrast, the three-point function $\langle \bar\psi\psi s\rangle$ is nontrivial. To leading order in $\epsilon$ it can be derived using the tree level cubic interaction and the star-triangle relation (\ref{star-triangle}),
\begin{equation}
\label{psi psi s leading}
\langle \bar\psi(x_1)\psi(x_2) s(x_3)\rangle = \tilde C_{\bar\psi\psi s}\,\frac{\gamma_\mu x_{13}^\mu
\gamma_\nu x_{32}^\nu}{|x_{12}|^{d-2\Delta_s}|x_{13}|^{d-2\Delta_\psi+1}|x_{23}|^{d-2\Delta_\psi+1}}\,,
\end{equation}
where the OPE coefficient is given by
\begin{equation}
\label{leading W psi psi s GNY}
\tilde C_{\bar\psi\psi s}=-g_1^\star\,
C_\psi \tilde C_s^\frac{1}{2}\pi^\frac{d}{2}A(1)V\left(\frac{d-1}{2},\frac{d-1}{2}\right)=-\sqrt{\frac{\epsilon}{N}}\,\left(1-\frac{3}{N} + {\cal O}\big(1/N^2\big)\right) \,.
\end{equation}
Note that, by definition of the OPE coefficient, all the fields are normalized so that their two-point functions have unit amplitude.

\section{CFT data for the critical Gross-Neveu model}
\label{sec: cft data in gross-neveu}

In this section we use the background field method \cite{Goldstone:1962es} to derive some CFT data for the critical Gross-Neveu model. Specifically, we calculate the three-point function $\langle \bar\psi(x_1)\psi(x_2)s(x_3)\rangle$ and the associated conformal effective vertex, up to  next-to-leading order in the $1/N$ expansion. In addition, we calculate the leading order OPE coefficient determined by the correlation function $\langle \bar\psi(x_1)\psi(x_2)s^2(x_3)\rangle$. Our calculations hold in general $2\leq d\leq 4$, and the final results agree with their counterparts obtained through the $\epsilon$-expansion of the critical Gross-Neveu-Yukawa model in the vicinity of $d=4$.   The latter serves as a check of our calculations, since the two critical models are equivalent \cite{ZinnJustin:1991yn}.

\subsection*{Evaluation of $\langle \bar\psi(x_1)\psi(x_2)s(x_3)\rangle$}
\label{leading psi psi s}

In what follows, the amplitudes of the two-point functions are normalized to unity, {\it i.e.,} we rescale the fields, 
\begin{equation}
\label{definition of normalization fields}
\psi \rightarrow \sqrt{C_\psi(1+A_\psi)}\,\psi\,,\qquad
s\rightarrow \sqrt{C_s(1+A_s)}\,s\,.
\end{equation}
The leading order three-point function for the normalized fields, which can be derived by using the tree level cubic vertex and the star-triangle relation,  reviewed in section \ref{sec:useful identities}, is,
\begin{equation}
\label{psi psi s leading}
\langle \bar\psi(x_1)\psi(x_2) s(x_3)\rangle = C_{\bar\psi\psi s}\,\frac{\gamma_\mu x_{13}^\mu
\gamma_\nu x_{32}^\nu}{|x_{12}|^{d-2\Delta_s}|x_{13}|^{d-2\Delta_\psi+1}|x_{23}|^{d-2\Delta_\psi+1}}~,
\end{equation}
where the leading order OPE coefficient is given by,
\begin{equation}
\label{leading W psi psi s GN}
C_{\bar\psi\psi s}=- \frac{1}{\sqrt{N}}\,
C_\psi C_s^\frac{1}{2}\pi^\frac{d}{2}A(1)V\left(\frac{d-1}{2},\frac{d-1}{2}\right)=\frac{-2^\frac{d}{2}}{\sqrt{N}(d-2)\pi^\frac{3}{4}}\,
\sqrt{-\frac{\sin \left(\frac{\pi  d}{2}\right) \Gamma \left(\frac{d-1}{2}\right)}{\Gamma \left(\frac{d}{2}-1\right)}}\,.
\end{equation}

Loops are associated with  higher order terms in $1/N$, which modify (\ref{psi psi s leading}). However, conformal invariance together with the discrete symmetry (\ref{parity like symmetry}) fixes the structure of the full three-point function. In our case, it takes the form,\footnote{Conformal symmetry uniquely determines the three point function of two fermion and one scalar field in terms of two possible structures \cite{Weinberg:2010fx,Goykhman:2018ihr}.
However, in our case the discrete symmetry (\ref{parity like symmetry}) eliminates one of them, and the end result is  (\ref{psi psi s subleading}). More generally, $\langle s^{2k+1}\psi\bar\psi\rangle$ and $\langle s^{2(k+1)}\psi\bar\psi\rangle$ for $k=0,1,2,\dots$ have the same form as (\ref{parity like symmetry}) and (\ref{psi psi s^2 general}) respectively, see also \cite{Goykhman:2020tsk} where $\langle s^2\psi\bar\psi \rangle$ was calculated at the next-to-leading order in $1/N$.}
\begin{equation}
\label{psi psi s subleading}
\langle \bar\psi(x_1)\psi(x_2) s(x_3)\rangle = C_{\bar\psi\psi s}\,(1+W_{\bar\psi\psi s})
\frac{\mu^{-2\gamma_\psi-\gamma_s}\;\gamma_\mu x_{13}^\mu \;
\gamma_\nu x_{32}^\nu}{|x_{13}|^{\Delta_s+\gamma_s+1}
|x_{32}|^{\Delta_s+\gamma_s+1}|x_{12}|^{2\Delta_\psi-\Delta_s+2\gamma_\psi-\gamma_s}}~,
\end{equation}
where the anomalous dimensions and $W_{\bar\psi\psi s}$ encompass loop corrections to the three-point function (\ref{psi psi s leading}). 

The anomalous dimensions $\gamma_s$ and $\gamma_\psi$ were evaluated in the previous section; in this section our goal is to determine $W_{\bar\psi\psi s}$. To this end, we need to dress the tree level diagram with $1/N$ corrections.  As we will see below, $W_{\bar\psi\psi s}$  receives contributions from the $1/N$ corrections to the amplitudes of external propagators,  as well as from the $1/N$ corrections to the cubic vertex itself. The former corrections were evaluated in the previous section ($A_s$ and $A_\psi$), so we only need to evaluate the latter terms.

The full cubic vertex of the effective action is non-local. By conformal invariance its form is fixed up to an overall constant prefactor, and can be represented diagrammatically as a conformal triangle \cite{Polyakov:1970xd},
\begin{center}
\begin{equation}
\label{definition of vertex Gamma}
  \begin{picture}(300,97) (23,3)
    \SetWidth{1.0}
    \SetColor{Black}
    \Line[arrow,arrowpos=0.5,arrowlength=5,arrowwidth=2,arrowinset=0.2](80,95)(52,47)
    \Line[](52,47)(108,47)
    \Line[arrow,arrowpos=0.5,arrowlength=5,arrowwidth=2,arrowinset=0.2](111,44)(83,96)
    \Vertex(81,97){4}
    \Vertex(52,47){4}
    \Vertex(111,47){4}
    \Text(69,107)[lb]{\scalebox{1}{$s(x_3)$}}
    \Text(20,35)[lb]{\scalebox{1}{$\bar\psi(x_1)$}}
    \Text(115,35)[lb]{\scalebox{01}{$\psi(x_2)$}}
    \Text(-40,66)[lb]{\scalebox{1}{$-\frac{\hat Z \mu^{2\gamma_\psi+\gamma_s}}{\sqrt{N}}~\times$}}
    \Text(102,73)[lb]{\scalebox{0.8}{$2\alpha$}}
    \Text(78,34)[lb]{\scalebox{0.8}{$2\beta$}}
    \Text(50,73)[lb]{\scalebox{0.8}{$2\alpha$}}
    \Text(120,65)[lb]{\scalebox{1}{$~=~-\frac{\hat Z}{\sqrt{N}}\,\int\int\int\,
\frac{\mu^{ 2\gamma_\psi+\gamma_s}\;x_{13}^\mu\gamma_\mu \; x_{32}^\nu\gamma_\nu}
{|x_{13}|^{2\alpha+1}|x_{32}|^{2\alpha+1}|x_{12}|^{2\beta}}\,\bar\psi(x_1)\psi(x_2)s(x_3)\,,$}}
  \end{picture}
  \end{equation}
\end{center}
where the unknown constant $\hat Z$ is closely related to $W_{\bar\psi\psi s}$, while $\alpha$ and $\beta$ are determined by noticing that each vertex, denoted by a solid dot in the above graph, should be conformal after attaching to it an appropriate full propagator (\ref{full psi propagator}) or (\ref{full s propagator}). Hence,
\begin{equation}
\alpha = \Delta_\psi-\frac{\gamma_s}{2}\,,\qquad \beta=\Delta_s-\gamma_\psi+\frac{\gamma_s}{2}~.
\end{equation}

We now focus on calculating  the only remaining unknown, $\hat Z$. By definition, it is determined by the following diagrammatic equation, which represents the perturbative expansion of the three-point function $\langle \bar\psi(x_1)\psi(x_2)s(x_3)\rangle$,
\begin{center}
\begin{equation}
  \begin{picture}(500,193) (6,3)
    \SetWidth{1.0}
    \SetColor{Black}
    \Text(62,157)[lb]{\scalebox{0.6}{$2\alpha$}}
    \Text(100,157)[lb]{\scalebox{0.6}{$2\alpha$}}
    \Text(82,132)[lb]{\scalebox{0.6}{$2\beta$}}
    \Line[](85,203)(85,167)
    \Vertex(85,167){4}
    \Line[arrow,arrowpos=0.5,arrowlength=5,arrowwidth=2,arrowinset=0.2](85,167)(65,140)
    \Vertex(65,140){4}
    \Line[arrow,arrowpos=0.5,arrowlength=5,arrowwidth=2,arrowinset=0.2](104,141)(85,167)
    \Vertex(104,141){4}
    \Line[](66,141)(104,141)
    \Line[arrow,arrowpos=0.5,arrowlength=5,arrowwidth=2,arrowinset=0.2](65,140)(38,122)
    \Line[arrow,arrowpos=0.5,arrowlength=5,arrowwidth=2,arrowinset=0.2](130,122)(104,141)
    \Text(5,160)[lb]{\scalebox{1}{$\hat Z\mu^{-2\gamma_\psi-\gamma_s}$}}
    %
    \Text(156,160)[lb]{\scalebox{1}{$=$}}
    \Line[](217,159)(217,205)
    \Line[arrow,arrowpos=0.5,arrowlength=5,arrowwidth=2,arrowinset=0.2](217,159)(181,122)
    \Line[arrow,arrowpos=0.5,arrowlength=5,arrowwidth=2,arrowinset=0.2](250,122)(217,159)
    \Vertex(217,159){4}
    %
    \Text(285,160)[lb]{\scalebox{1}{$+$}}
    \Text(300,157)[lb]{\scalebox{1}{$\frac{2\gamma_\psi+\gamma_s}{\delta}~\times$}}
    \Line[](387,159)(387,205)
    \Line[arrow,arrowpos=0.5,arrowlength=5,arrowwidth=2,arrowinset=0.2](387,159)(352,122)
    \Line[arrow,arrowpos=0.5,arrowlength=5,arrowwidth=2,arrowinset=0.2](421,122)(387,159)
    \Vertex(387,159){4}
    %
    \Text(56,42)[lb]{\scalebox{1}{$+~\frac{C_\psi^2 C_s\mu^{-\delta}}{N}~\times$}}
    \Text(136,30)[lb]{\scalebox{0.6}{$2\Delta_\psi$}}
    \Text(188,30)[lb]{\scalebox{0.6}{$2\Delta_\psi$}}
    \Text(156,13)[lb]{\scalebox{0.6}{$2\Delta_s+\delta$}}
    \Line[](167,88)(167,51)
    \Vertex(167,51){4}
    \Line[arrow,arrowpos=0.5,arrowlength=5,arrowwidth=2,arrowinset=0.2](167,51)(148,23)
    \Line[arrow,arrowpos=0.5,arrowlength=5,arrowwidth=2,arrowinset=0.2](187,23)(167,51)
    \Line[](148,23)(187,23)
    \Vertex(148,23){4}
    \Vertex(187,23){4}
    \Line[arrow,arrowpos=0.5,arrowlength=5,arrowwidth=2,arrowinset=0.2](148,23)(120,4)
    \Line[arrow,arrowpos=0.5,arrowlength=5,arrowwidth=2,arrowinset=0.2](213,4)(187,23)
    %
    \Text(246,42)[lb]{\scalebox{1}{$+~\frac{C_\psi^4 C_s^2\mu^{-2\delta}}{N}~\times$}}
    \Text(315,30)[lb]{\scalebox{0.6}{$2\Delta_s+\delta$}}
    \Text(393,30)[lb]{\scalebox{0.6}{$2\Delta_s+\delta$}}
    \Text(362,13)[lb]{\scalebox{0.6}{$2\Delta_\psi$}}
    \Text(362,31)[lb]{\scalebox{0.6}{$2\Delta_\psi$}}
    \Text(342,57)[lb]{\scalebox{0.6}{$2\Delta_\psi$}}
    \Text(380,57)[lb]{\scalebox{0.6}{$2\Delta_\psi$}}
    \Line[](367,88)(367,73)
    \Vertex(367,73){4}
    \Line[arrow,arrowpos=0.35,arrowlength=5,arrowwidth=2,arrowinset=0.2](367,73)(340,23)
    \Line[arrow,arrowpos=0.65,arrowlength=5,arrowwidth=2,arrowinset=0.2](391,23)(367,73)
    \Line[arrow,arrowpos=0.5,arrowlength=5,arrowwidth=2,arrowinset=0.2](350,42)(382,42)
    \Vertex(350,42){4}
    \Vertex(382,42){4}
    \Line[arrow,arrowpos=0.5,arrowlength=5,arrowwidth=2,arrowinset=0.2](391,23)(340,23)
    \Vertex(391,23){4}
    \Vertex(340,23){4}
    \Line[arrow,arrowpos=0.5,arrowlength=5,arrowwidth=2,arrowinset=0.2](340,23)(322,4)
    \Line[arrow,arrowpos=0.5,arrowlength=5,arrowwidth=2,arrowinset=0.2](410,4)(391,23)
  \end{picture}
   \label{Dressed vertex diagram}
\end{equation}
\end{center}
where the external legs are not amputated; 
 they correspond to the {\it full} propagators (\ref{full psi propagator}) and (\ref{full s propagator}).

The first graph on the right-hand side represents the
cubic interaction of (\ref{starting action HS}).
The rest of the terms reveal the $1/N$ terms.
Specifically, the two diagrams on the second line of (\ref{Dressed vertex diagram}) account for the
loop corrections to the leading order cubic vertex.
In fact, the last diagram 
is trivial, because it contains a vanishing $\langle sss\rangle$ sub-graph,
discussed in section \ref{GNYsub}.
On the other hand, the first diagram on the second line of (\ref{Dressed vertex diagram})
results in a non-trivial next-to-leading order contribution to the cubic vertex.~\footnote{
The two last graphs of (\ref{Dressed vertex diagram}) are analogous to the corresponding vertex diagrams in the $O(N)$ vector model. However, unlike the case studied here, in the critical $O(N)$ vector model both diagrams give rise to a non-trivial contribution, see, e.g., \cite{Goykhman:2019kcj} for a recent discussion. In appendix~\ref{app:vertex correction in ON vector model}
we recover $\hat Z$ in the vector model using the background field method.}
This diagram diverges and can be regularized by a small shift $\delta$ in the scaling dimension of $s$. This is the same regulator $\delta$ used  in section \ref{sec:corrections to propagators} to calculate the
 $1/N$ corrections to the propagators of the fields $\psi$ and $s$.

While the couplings are not renormalized at the fixed point, there is still renormalization of the fields. In particular, the second diagram on the  right-hand side of (\ref{Dressed vertex diagram}) is associated with the counterterm (\ref{vertex counterterm}) induced by  wave function renormalization. In the limit $\delta\rightarrow 0$, this counterterm cancels the divergences of the loop diagrams. 

The diagrammatic equation (\ref{Dressed vertex diagram}) simplifies if the external propagator of the auxiliary field $s$ is replaced with a constant background $\bar s$. At the level of the path integral, this is tantamount to decomposing the Hubbard-Stratonovich field into a fluctuating component $s$ and a constant background $\bar s$, {\it i.e.,} $s\rightarrow \bar s+s$. The resulting diagrammatic equation represents a linear (in $\bar s$) correction to the two-point function $\langle\bar\psi(x_1)\psi(x_2)\rangle_{\bar s}$ in the presence of a constant background, 
\begin{center}
\begin{equation}
  \begin{picture}(500,150) (6,0)
    \SetWidth{1.0}
    \SetColor{Black}
    \Text(82,177)[lb]{\scalebox{1}{$\bar s$}}
    \Text(235,168)[lb]{\scalebox{1}{$\bar s$}}
    \Text(385,168)[lb]{\scalebox{1}{$\bar s$}}
     \Text(145,70)[lb]{\scalebox{1}{$\bar s$}}
     \Text(365,90)[lb]{\scalebox{1}{$\bar s$}}
    \Text(62,157)[lb]{\scalebox{0.6}{$2\alpha$}}
    \Text(100,157)[lb]{\scalebox{0.6}{$2\alpha$}}
    \Text(82,132)[lb]{\scalebox{0.6}{$2\beta$}}
    \Vertex(85,167){4}
    \Line[arrow,arrowpos=0.5,arrowlength=5,arrowwidth=2,arrowinset=0.2](85,167)(65,140)
    \Vertex(65,140){4}
    \Line[arrow,arrowpos=0.5,arrowlength=5,arrowwidth=2,arrowinset=0.2](104,141)(85,167)
    \Vertex(104,141){4}
    \Line[](66,141)(104,141)
    \Line[arrow,arrowpos=0.5,arrowlength=5,arrowwidth=2,arrowinset=0.2](65,140)(38,122)
    \Line[arrow,arrowpos=0.5,arrowlength=5,arrowwidth=2,arrowinset=0.2](130,122)(104,141)
    \Vertex(38,122){2}
    \Vertex(130,123){2}
    \Text(5,145)[lb]{\scalebox{1}{$\hat Z\mu^{-2\gamma_\psi+\gamma_s}$}}
    \Text(50,120)[lb]{\scalebox{0.6}{$2(\Delta_\psi+\gamma_\psi)$}} 
    \Text(123,130)[lb]{\scalebox{0.6}{$2(\Delta_\psi+\gamma_\psi)$}} 
    \Text(156,145)[lb]{\scalebox{1}{$=\mu^{-4\gamma_\psi}~\times$}}
    \Line[arrow,arrowpos=0.5,arrowlength=5,arrowwidth=2,arrowinset=0.2](237,159)(201,122)
    \Line[arrow,arrowpos=0.5,arrowlength=5,arrowwidth=2,arrowinset=0.2](270,122)(237,159)
    \Vertex(237,159){4}
    \Vertex(201,122){2}
    \Vertex(270,122){2}
    \Text(215,125)[lb]{\scalebox{0.6}{$2(\Delta_\psi+\gamma_\psi)$}} 
    \Text(267,130)[lb]{\scalebox{0.6}{$2(\Delta_\psi+\gamma_\psi)$}} 
    \Text(285,145)[lb]{\scalebox{1}{$+$}}
    \Text(300,142)[lb]{\scalebox{1}{$\frac{2\gamma_\psi+\gamma_s}{\delta}~\times$}}
    \Line[arrow,arrowpos=0.5,arrowlength=5,arrowwidth=2,arrowinset=0.2](387,159)(352,122)
    \Line[arrow,arrowpos=0.5,arrowlength=5,arrowwidth=2,arrowinset=0.2](421,122)(387,159)
    \Vertex(352,122){2}
    \Vertex(421,122){2}
    \Vertex(387,159){4}
    \Text(365,125)[lb]{\scalebox{0.6}{$2\Delta_\psi$}} 
    \Text(418,130)[lb]{\scalebox{0.6}{$2\Delta_\psi$}} 
    \Text(36,22)[lb]{\scalebox{1}{$+~\frac{C_\psi^2 C_s\mu^{-\delta}}{N}~\times$}}
    \Text(116,40)[lb]{\scalebox{0.6}{$2\Delta_\psi$}}
    \Text(168,40)[lb]{\scalebox{0.6}{$2\Delta_\psi$}}
    \Text(136,23)[lb]{\scalebox{0.6}{$2\Delta_s+\delta$}}
    \Vertex(147,61){4}
    \Line[arrow,arrowpos=0.5,arrowlength=5,arrowwidth=2,arrowinset=0.2](147,61)(128,33)
    \Line[arrow,arrowpos=0.5,arrowlength=5,arrowwidth=2,arrowinset=0.2](167,33)(147,61)
    \Line[](128,33)(167,33)
    \Vertex(128,33){4}
    \Vertex(167,33){4}
    \Line[arrow,arrowpos=0.5,arrowlength=5,arrowwidth=2,arrowinset=0.2](128,33)(100,14)
    \Line[arrow,arrowpos=0.5,arrowlength=5,arrowwidth=2,arrowinset=0.2](193,14)(167,33)
    \Vertex(100,14){2}
    \Vertex(193,14){2}
    \Text(115,14)[lb]{\scalebox{0.6}{$2\Delta_\psi$}} 
    \Text(190,20)[lb]{\scalebox{0.6}{$2\Delta_\psi$}} 
    \Text(240,22)[lb]{\scalebox{1}{$+~\frac{C_\psi^4 C_s^2\mu^{-2\delta}}{N}~\times$}}
    \Text(315,40)[lb]{\scalebox{0.6}{$2\Delta_s+\delta$}}
    \Text(393,40)[lb]{\scalebox{0.6}{$2\Delta_s+\delta$}}
    \Text(362,23)[lb]{\scalebox{0.6}{$2\Delta_\psi$}}
    \Text(362,41)[lb]{\scalebox{0.6}{$2\Delta_\psi$}}
    \Text(342,67)[lb]{\scalebox{0.6}{$2\Delta_\psi$}}
    \Text(380,67)[lb]{\scalebox{0.6}{$2\Delta_\psi$}}
    \Vertex(367,83){4}
    \Line[arrow,arrowpos=0.35,arrowlength=5,arrowwidth=2,arrowinset=0.2](367,83)(340,33)
    \Line[arrow,arrowpos=0.65,arrowlength=5,arrowwidth=2,arrowinset=0.2](391,33)(367,83)
    \Line[arrow,arrowpos=0.5,arrowlength=5,arrowwidth=2,arrowinset=0.2](350,52)(382,52)
    \Vertex(350,52){4}
    \Vertex(382,52){4}
    \Line[arrow,arrowpos=0.5,arrowlength=5,arrowwidth=2,arrowinset=0.2](391,33)(340,33)
    \Vertex(391,33){4}
    \Vertex(340,33){4}
    \Line[arrow,arrowpos=0.5,arrowlength=5,arrowwidth=2,arrowinset=0.2](340,33)(322,14)
    \Line[arrow,arrowpos=0.5,arrowlength=5,arrowwidth=2,arrowinset=0.2](410,14)(391,33)
    \Vertex(322,14){2}
    \Vertex(410,14){2}
    \Text(335,15)[lb]{\scalebox{0.6}{$2\Delta_\psi$}} 
    \Text(412,20)[lb]{\scalebox{0.6}{$2\Delta_\psi$}} 
  \end{picture}
   \label{Dressed fermion propagator in s background}
\end{equation}
\end{center}
Note that the amplitudes of the external fermionic propagators (\ref{full psi propagator}) cancel  on both sides of this equation, and we ignore them in what follows.

Each diagram in (\ref{Dressed fermion propagator in s background}) can be calculated by repeatedly applying 
the fermion-fermion and fermion-scalar propagator merging relations reviewed in section \ref{sec:useful identities}. The calculation of each diagram starts by integrating over the insertion point of the vertex with a constant background $\bar s$. In particular, the left-hand side  of (\ref{Dressed fermion propagator in s background}) takes the form,
\begin{align}
\text{l.h.s. of (\ref{Dressed fermion propagator in s background})} &=
\pi^\frac{3d}{2}\,\hat Z\,A(1+\gamma_s)V\left(\frac{d-1-\gamma_s}{2},\frac{d-1-\gamma_s}{2}\right)\\
&\times A\left(\frac{d-\gamma_s}{2}-\gamma_\psi\right)V\left(\frac{d-1}{2}+\gamma_\psi,\frac{1+\gamma_s}{2}\right)
\notag\\
&\times A\left(1-\gamma_\psi+\frac{\gamma_s}{2}\right)V\left(\frac{d-1-\gamma_s}{2},\frac{d-1}{2}+\gamma_\psi\right)\,
\frac{\mu^{-2\gamma_\psi+\gamma_s} \bar s}{|x_{12}|^{d-2+2\gamma_\psi-\gamma_s}}~.
\notag
\end{align}
Substituting
\begin{equation} \label{ZZ0}
\hat Z = Z_0 (1+\delta Z)\,,
\end{equation}
where $\delta Z \sim {\cal O}(1/N)$, and expanding in the small parameters $\gamma_s, \gamma_\psi\sim  {\cal O}(1/N)$, yields
\begin{align}
\label{psi psi bar s triangle}
\text{l.h.s. of (\ref{Dressed fermion propagator in s background})}  &=
\frac{8 \pi ^{\frac{3 d}{2}} \, Z_0}{(d-2)^2\Gamma\big(\frac{d}{2}\big)^3(\gamma_s+2\gamma_\psi)}\,
\frac{\bar s}{|x_{12}|^{d-2}}\\
&\times \left(1+\delta Z +{3\gamma_s-2\gamma_\psi\over d-2}-(2\gamma_\psi-\gamma_s)\,\log(|x_{12}|\mu)\right)\,.\notag
\end{align}

On the other hand, this expression is equal to the sum of the three diagrams on the right-hand side of 
(\ref{Dressed fermion propagator in s background}), which we denote by,~\footnote{Recall that the last  diagram in (\ref{Dressed fermion propagator in s background}) vanishes, because it has a subdiagram $\langle sss\rangle$ which does not respect $s\to -s$ symmetry.}
\begin{equation}
\label{psi psi bar s triangle in terms of sum of vs}
\text{r.h.s. of (\ref{Dressed fermion propagator in s background})} = v_\text{tree} + v_{\textrm{c.t.}}+v_\text{loop}(\delta)\,.
\end{equation}
The tree level term is given by,
\begin{equation}
v_\text{tree} = -\pi^\frac{d}{2}\,A(1-2\gamma_\psi)V\left(\frac{d-1}{2}+\gamma_\psi,
\frac{d-1}{2}+\gamma_\psi\right)\,\frac{\mu^{-4\gamma_\psi}\bar s}{|x_{12}|^{d-2+4\gamma_\psi}}~.
\end{equation}
It contributes to the leading and subleading orders in $1/N$,
\begin{equation}
\label{v0}
v_\text{tree}  = -\frac{2 \pi ^{d/2}}{(d-2) \Gamma \left(\frac{d}{2}\right)}
\,\left(1+\frac{1}{N}\,\frac{4 \sin \left(\frac{\pi  d}{2}\right) \Gamma (d-1)}{\pi  d \Gamma \left(\frac{d}{2}\right)^2}
-4\gamma_\psi \log(|x_{12}|\mu)\right) 
\,\frac{\bar s}{|x_{12}|^{d-2}}+\mathcal{O}(1/N^2)~.
\end{equation}
The counterterm  satisfies
\begin{equation}
\label{vct}
v_{\textrm{c.t.}} = -\frac{2 \pi ^{d/2}}{(d-2) \Gamma \left(\frac{d}{2}\right)}
\,\frac{2\gamma_\psi+\gamma_s}{\delta}\,\frac{\bar s}{|x_{12}|^{d-2}}~.
\end{equation}
Finally, the loop diagram takes the form
\begin{align}
v_\text{loop}(\delta) &= \frac{C_\psi^2 C_s }{N}\,\frac{\bar s}{(\mu |x_{12}|)^{d-2+\delta}}\,\pi^\frac{3d}{2}\,
A(1)V\left(\frac{d-1}{2},\frac{d-1}{2}\right)
A\left(\frac{d+\delta}{2}\right)\notag\\
&\times V\left(\frac{d-1}{2},\frac{1-\delta}{2}\right)
A\left(1-\frac{\delta}{2}\right)V\left(\frac{d-1+\delta}{2},\frac{d-1}{2}\right)~.
\end{align}
Expanding  around $\delta = 0$ yields,
\begin{align}
\label{v12}
v_\text{loop}(\delta) &=- \frac{2 \pi ^{d/2}}{(d-2) \Gamma \left(\frac{d}{2}\right)}
\,\frac{\bar s}{|x_{12}|^{d-2}}\\
&\times\,\left(-\frac{2\gamma_\psi+\gamma_s}{\delta}+(2\gamma_\psi+\gamma_s)\,\log(|x_{12}|\mu)
+\frac{1}{N}\,\frac{2^{d-1} \sin \left(\frac{\pi  d}{2}\right) \Gamma \left(\frac{d-1}{2}\right)}{\pi ^{3/2} (d-2) \Gamma \left(\frac{d}{2}\right)}\right)\,.\notag
\end{align}
As expected, the sum of the three terms (\ref{v0}), (\ref{vct}) and (\ref{v12}) is finite in the limit $\delta\rightarrow 0$
\begin{align}
\text{r.h.s. of (\ref{Dressed fermion propagator in s background})} &= v_\text{tree} + v_{\textrm{c.t.}}+v_\text{loop}(\delta) \underset{\delta\to 0}{\to}
-\frac{2 \pi ^{d/2}}{(d-2) \Gamma \left(\frac{d}{2}\right)}\,
\frac{\bar s}{|x_{12}|^{d-2}}\label{v total}\\
&\times \left(1+\frac{1}{N}\,\frac{2 (3 d-4) \sin \left(\frac{\pi  d}{2}\right) \Gamma (d-1)}{\pi  (d-2) d \Gamma \left(\frac{d}{2}\right)^2}-(2\gamma_\psi-\gamma_s)\,\log(|x_{12}|\mu)\right)\,.\notag
\end{align}
Comparing (\ref{psi psi bar s triangle}) to
(\ref{v total}), we get $\hat Z$ (\ref{ZZ0}),
\begin{align}
Z_0 &
=-\frac{(d-2)\Gamma \big(\frac{d}{2}\big)^2}{4\pi^d} \,(2\gamma_\psi+\gamma_s)\,,\\
\delta Z &= - {8\Gamma(d) \sin\big({\pi d\over 2}\big) \over Nd(d-2)\pi \Gamma \big(\frac{d}{2}\big)^2} 
=-\frac{2}{d-2}\,\gamma_s\,.\label{delta Z}
\end{align}
The same result was recently obtained in \cite{Manashov:2017rrx}, using a different method.

Finally, calculating the three-point function $\langle \bar\psi(x_1)\psi(x_2)s(x_3)\rangle$  simply involves attaching the full  propagators (\ref{full psi propagator}), (\ref{full s propagator}) to the effective cubic vertex (\ref{definition of vertex Gamma}), and integrating over the internal vertices through the use of the star-triangle relation (\ref{star-triangle}). After normalizing the external fields as in (\ref{definition of normalization fields}), the final result reads,
\begin{equation}
\label{psi psi s correction}
\langle \bar\psi(x_1)\psi(x_2) s(x_3)\rangle = -\frac{C_\psi(1{+}A_\psi) \sqrt{C_s(1{+}A_s)}}{\sqrt{N}} \,
\frac{ \hat Z\, U \mu^{-2\gamma_\psi-\gamma_s}\;\gamma_\mu x_{13}^\mu \;
\gamma_\nu x_{32}^\nu}{|x_{13}|^{\Delta_s+\gamma_s+1}
|x_{32}|^{\Delta_s+\gamma_s+1}|x_{12}|^{2\Delta_\psi-\Delta_s+2\gamma_\psi-\gamma_s}}\,,
\end{equation}
where 
\begin{align}
U &= -\pi^\frac{3d}{2}\,A(1+\gamma_s)\,V\left(\frac{d-1-\gamma_s}{2},\frac{d-1-\gamma_s}{2}\right)\,
A\left(\frac{d-\gamma_s}{2}-\gamma_\psi\right)\,
\notag\\
&\times V\left(\frac{1+\gamma_s}{2},\frac{d-1}{2}+\gamma_\psi\right)
A\left(1-\gamma_\psi+\frac{\gamma_s}{2}\right)\,
V\left(\frac{d-1-\gamma_s}{2},\frac{d-1}{2}+\gamma_\psi\right)\notag\\
&=-\frac{8 \pi ^{\frac{3 d}{2}} }{(d-2)^2\Gamma\big(\frac{d}{2}\big)^3(\gamma_s+2\gamma_\psi)}
\Big(1+\delta U +{\cal O}(1/N^2)\ \Big) ~, \ \ \ \ \ \ \ \  \delta U={3\gamma_s-2\gamma_\psi\over d-2}~.
\end{align}
In the last line we expanded in the small anomalous dimensions. Expanding the right-hand side of (\ref{psi psi s correction}), up to  next-to-leading order in $1/N$, and comparing to (\ref{psi psi s subleading}),
we deduce that the leading term reproduces (\ref{leading W psi psi s GN}), while the next-to-leading order correction is given by, \footnote{The analogous OPE coefficient in the $O(N)$ vector model was recently derived in \cite{Goykhman:2019kcj}.}
\begin{equation}
\label{W psi psi s result}
W_{\bar\psi\psi s} = A_\psi+\frac{A_s}{2} +\delta Z + \delta U
=\frac{2 (d-1) \left((d-2) H_{d-2}+\pi  (d-2) \cot \left(\frac{\pi  d}{2}\right)-2\right)}{(d-2)^2}\,
\gamma_\psi~.
\end{equation}

\begin{figure}[t]
\begin{center}
\includegraphics[width=220pt]{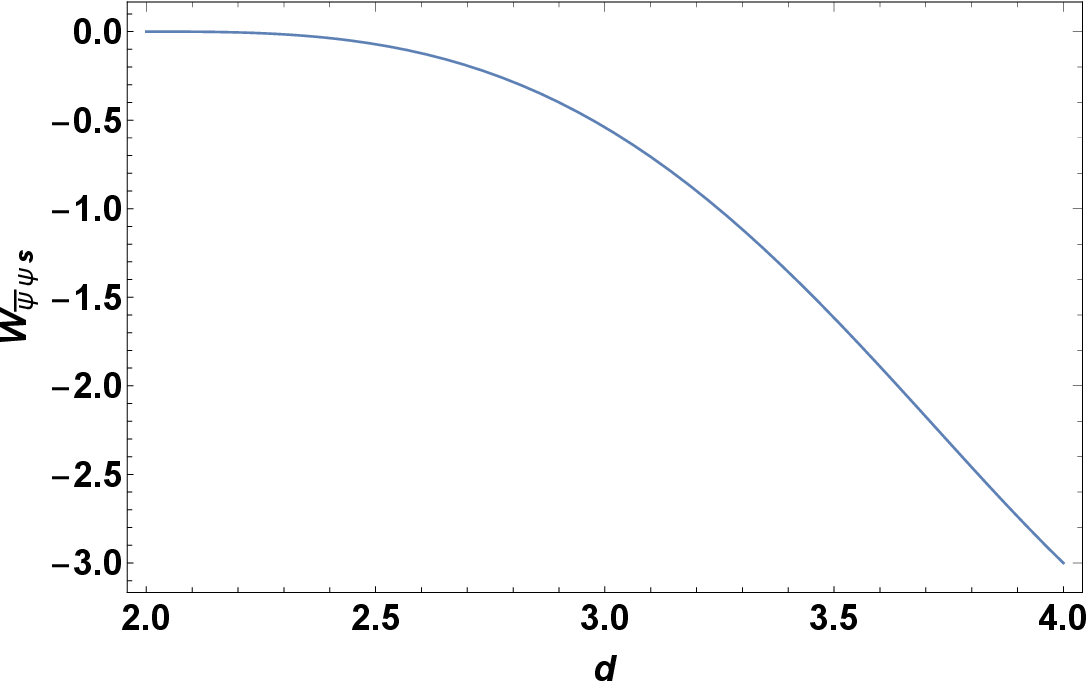}
\end{center}
\caption{The $1/N$ correction to the OPE coefficient, $W_{\bar\psi\psi s}$ (defined in Eq.~\ref{psi psi s subleading}), as a function of space-time dimension $2\leq d \leq4$. The end points of the plot satisfy $W_{\bar\psi\psi s} (d = 2) = 0$ and $ W_{\bar\psi\psi s} (d = 4) = -3$.}
\label{fig:w_psi psi s}
\end{figure}

It is instructive to check our result for $W_{\bar\psi\psi s}$. To this end, we substitute (\ref{leading W psi psi s GN}) and (\ref{W psi psi s result}) into (\ref{psi psi s subleading}), and expand the three-point function $\langle \bar\psi \psi s\rangle$ in $\epsilon=4-d$.  The leading order term in $\epsilon$ matches (\ref{leading W psi psi s GNY}), in accord with the equivalence between the critical Gross-Neveu and Gross-Neveu-Yukawa models. Moreover, $W_{\bar\psi\psi s}$ vanishes in $d=2$, and therefore the OPE coefficient equals $-1$, up to  order $1/N$. This result is expected, because the critical GN model is free in two dimensions and $\langle \bar\psi \psi s\rangle \sim \langle \bar\psi \psi (\bar\psi\psi)\rangle$, since $s \sim \bar\psi\psi/\sqrt{N}$, by  the equations of motion. In particular, it follows from the standard Wick contraction applied to a free field theory (with properly normalized fields) that the OPE coefficient is indeed $-1$, to all orders in $1/N$. Figure \ref{fig:w_psi psi s} displays $W_{\bar\psi\psi s}$ for dimensions $2\leq d \leq 4$.

\subsection*{Evaluation of  $\langle \bar\psi(x_1)\psi(x_2)s^2(x_3)\rangle$}
\label{sec:psi psi s^2}

Let us close this section by deriving the three-point function $\langle \bar\psi\psi s^2\rangle$
between the Dirac fields and the composite operator $s^2$. To leading order in the $1/N$ expansion, this correlator is not affected by the anomalous dimensions of $\psi$ and $s^2$,
\begin{equation}
\label{psi psi s^2 general}
\langle s^2 (x_1)\psi(x_2)\bar\psi(x_3)\rangle = C_{s^2\bar\psi\psi}\,
\frac{x_{23}^\mu \gamma_\mu}{|x_{12}|^2|x_{13}|^2 |x_{23}|^{d-2}}\, \Big( 1+ {\cal O}(1/N)\Big).
\end{equation}
On the other hand, it is determined by the diagram
\begin{center}
  \begin{picture}(104,75) (57,3)
    \SetWidth{1.0}
    \SetColor{Black}
    \CBox(114,58)(126,70){Black}{Black}
    \Line[](114,58)(96,28)
    \Line[](126,58)(144,28)
    \Line[arrow,arrowpos=0.5,arrowlength=5,arrowwidth=2,arrowinset=0.2](144,28)(96,28)
    \Line[arrow,arrowpos=0.5,arrowlength=5,arrowwidth=2,arrowinset=0.2](96,28)(66,10)
    \Line[arrow,arrowpos=0.5,arrowlength=5,arrowwidth=2,arrowinset=0.2](174,10)(144,28)
    \Vertex(96,28){4}
    \Vertex(144,28){4}
    \Vertex(66,10){2}
    \Vertex(174,10){2}
    \Text(0,22)[lb]{\scalebox{1}{$C_s^2C_\psi^3\times$}}
    \Text(117,75)[lb]{$x_1$}
    \Text(54,-2)[lb]{$x_2$}
    \Text(180,-2)[lb]{$x_3$}
    \Text(55,22)[lb]{$d-1$}
    \Text(110,15)[lb]{$d-1$}
    \Text(162,22)[lb]{$d-1$}
    \Text(95,46)[lb]{$2$}
    \Text(140,46)[lb]{$2$}
  \end{picture}
\end{center}
where insertion of $s^2$ is denoted by a black square.

Using the star-triangle relation (\ref{star-triangle}) to integrate over the internal vertices, and normalizing $s^2$ and the Dirac fields
according to $s^2\rightarrow \sqrt{2} \,C_s\,s^2$ and (\ref{definition of normalization fields}), respectively, yields
\begin{equation}
\label{C s^2 bar psi psi}
C_{s^2\bar\psi\psi} = \frac{1}{N}
\frac{2^{d-\frac{3}{2}} \sin \left(\frac{\pi  d}{2}\right) \Gamma \left(\frac{d-1}{2}\right)}{\pi ^{3/2} \Gamma \left(\frac{d}{2}\right)}\,.
\end{equation}
This result is in full agreement with the GNY model.~\footnote{In the GNY model $\langle \bar\psi\psi s^2\rangle$ is determined by a similar diagram. To explicitly verify that it matches the GN counterpart, it is sufficient to use the leading order relation $(g_1^{\star})^2\,\tilde C_s/C_s=1+{\cal O}(\epsilon,1/N)$,
see section~\ref{GNYsub}.} Moreover, the three point function (\ref{psi psi s^2 general}) respects discrete symmetry (\ref{parity like symmetry}). Indeed, 
\begin{equation}
\langle s(x_1)^2 \psi(x_2)\bar\psi(x_3)\rangle\rightarrow
-\gamma^\mu \langle s(x_1)^2 \psi(x_2)\bar\psi(x_3)\rangle \gamma^\mu\quad\Rightarrow\quad
C_{s^2\bar\psi\psi}\rightarrow C_{s^2\bar\psi\psi}\,.
\end{equation}

\section{Discussion}
\label{sec:discussion}
In this paper we used the background field method \cite{Goldstone:1962es} as a simple way to calculate  conformal cubic vertices in  critical ${\cal O}(N)$ vector models, up to next-to-leading order in $1/N$. To the best of our knowledge, this method of calculation has not been previously employed in large-$N$ conformal perturbation theory. 

We explicitly studied two critical vector models: the $O(N)$ sextic $\phi^6$ model in dimension $d=3$,  and the $U(n)$ Gross-Neveu model in dimensions $2\leq d\leq 4$. For the $\phi^6$ model, we derived the leading order anomalous dimensions, (\ref{gamma rho}) and (\ref{gamma rho n}), of the composite operators $\rho^n\sim (\phi^2)^n$, as well as  the OPE coefficients associated with the three-point functions $\langle \phi^2\phi^2\phi^2\rangle$ and $\langle \phi \phi \phi^2\rangle$, up to the next-to-leading order in $1/N$,  see  (\ref{next to-leading normalized rho rho rho}) and  (\ref{phi phi rho general}).
For the critical Gross-Neveu model, we derived the leading order and next-to-leading order OPE coefficient of the three-point functions $\langle \bar\psi\psi s\rangle$ and
$\langle \bar\psi\psi s^2\rangle$, see  (\ref{leading W psi psi s GN}), (\ref{W psi psi s result}) and (\ref{C s^2 bar psi psi}).
Our results agree with the literature \cite{Pisarski:1982vz,Manashov:2017rrx} and extend it in several ways.

There are a number of motivations for studying critical vector models. 
The Gross-Neveu  model, specifically, is asymptotically safe despite being non-renormalizable in the vicinity of the Gaussian fixed point above two dimensions. Its RG flow is  therefore completely determined by a finite number of parameters, and its UV fixed point has been extensively studied in the large-$N$ limit \cite{ZinnJustin:1991yn,Gracey:1990wi,Gracey:1992cp,Derkachov:1993uw,Vasiliev:1992wr,Vasiliev:1993pi,Gracey:1993kb,Gracey:1993kc,Manashov:2016uam,Manashov:2017rrx,Diab:2016spb}. In particular, it was shown that the UV CFT of the model is equivalent to an IR CFT of the renormalizable Gross-Neveu-Yukawa model \cite{ZinnJustin:1991yn}. From this perspective, the Gross-Neveu model provides a  laboratory where theoretical questions about the structure of RG flows and mechanisms of UV completion can be explicitly addressed. In two dimensions it presents a simple example of an asymptotically free field theory \cite{Gross:1974jv}.

In contrast, the $\phi^6$ vector model is manifestly renormalizable in three dimensions, yet it also possesses a UV fixed point which is interesting in its own way.  It turns out that the UV physics of the model is closely related to the study of big crunch singularities in asymptotically $AdS_4$ spacetimes \cite{Craps:2009qc,Smolkin:2012er}, see also \cite{Craps:2007ch}. In the latter context, one studies a marginal triple trace deformation of ABJM theory \cite{ABJM}, where the deformation corresponds to adding a potential which is unbounded from below. The UV fixed point of the deformed ABJM theory in the limit of zero 't Hooft coupling is described by the critical $\phi^6$ model studied in this work. 
It would be interesting to systematically explore the effects of the non-perturbative instability in $1/N$ \cite{Bardeen:1983rv} on the CFT data of the critical $\phi^6$ model and understand the cosmological implications of our findings in the holographic context studied in \cite{Craps:2009qc,Smolkin:2012er}.

 It is tempting to study the critical $\phi^6$ vector model away from $d=3$. By continuity, the UV fixed point does not disappear as we move towards other dimensions, provided that $N$ is sufficiently large. Thus, we expect to generate a family of CFTs parameterized by $d$, and a natural question is whether there is a range for $d$ such that the corresponding CFTs are non-perturbatively stable.

Finally, it turns out that  critical vector models exhibit peculiar behavior when coupled to a thermal bath. For instance, it was shown recently that there are $O(N)$ theories which have some of their internal symmetries broken at arbitrary finite temperature \cite{Chai:2020zgq}. The results of this paper might be of help in understanding this behavior in three dimensions.

\section*{Acknowledgements} \noindent We thank Noam Chai, Soumyadeep Chaudhuri, Eliezer Rabinovici and Ritam Sinha for helpful discussions. The work of MG and MS is partially supported by the Binational Science Foundation (grant No. 2016186), the Israeli Science Foundation Center of Excellence (grant No. 2289/18) and by the Quantum Universe I-CORE program of the Israel Planning and Budgeting Committee (grant No. 1937/12). The work of VR is supported by NSF grant PHY-1911298, as well as the Sivian Fund of the IAS.

\appendix

\section{Interaction vertex in the critical $\phi^4$  model}
\label{app:vertex correction in ON vector model}

In this appendix we derive the effective interaction vertex in the critical $\phi^4$ vector model in $2\leq d \leq 6$ dimensions. The $\phi^4$ vector model has an IR fixed point in dimensions  $2\leq d \leq 4$, and a UV fixed point in dimensions $4<d \leq6$. For a recent discussion of this model see \cite{Goykhman:2019kcj} and references therein, including \cite{Vasiliev:1981yc,Vasiliev:1981dg,Lang:1993ct,Petkou:1995vu,Petkou:1994ad,Derkachov:1997ch,Leonhardt:2003du,Gracey:2018ame,Alday:2019clp,Giombi:2019upv}.
Our goal is to demonstrate that the background field method, used in section~\ref{sec: cft data in gross-neveu}
to calculate the effective $\bar\psi\psi s$
vertex in the Gross-Neveu model, is also applicable to the $O(N)$
vector model. Our result for the cubic vertex in the $\phi^4$ model fully agrees with \cite{Derkachov:1997ch}, where it was evaluated using a different method. Since, to the best of our knowledge, the background field method has not been used before in the context of  large-$N$ conformal perturbation theory, we felt some researchers may find the current discussion useful. The discussion in this appendix parallels the discussion in section~\ref{sec: cft data in gross-neveu}.

The critical $\phi^4$ vector model is described by the action,
\begin{equation}
\label{action of vector model}
S = \int d^dx \, \left(\frac{1}{2}\, (\partial\phi) ^ 2 + \frac{1}{\sqrt{N}}\,s\,\phi^2 + \frac{2\gamma_\phi
+\gamma_s}{{\sqrt{N}}\, \delta}\, s\phi^2 \right)\,,
\end{equation}
where $\phi$ is a real  $N$-component multiplet of scalar fields in the fundamental
representation  of $O(N)$, $s$ is the auxiliary Hubbard-Stratonovich field, $\delta$ is a UV regulator, and
$\gamma_{\phi,s}$ are the anomalous dimensions associated with the loop corrections to the large-$N$ scaling dimensions $\Delta_\phi = d/2-1$ and  $\Delta_s = 2$ of $\phi$ and $s$, respectively. Note that the Hubbard-Stratonovich field is dynamical; its propagator is built from an infinite series of $\phi$ bubbles.~\footnote{The effective action for $s$ includes an infinite tower of non-local $n$-vertices suppressed by $1/N^{\frac{n}{2}-1}$.}

The UV divergences of the theory are regulated by adding a small shift to the scaling dimension of the internal $s$-lines in the Feynman graphs, {\it i.e.,} $2\Delta_s\to2\Delta_s+\delta$. In particular, the last term in (\ref{action of vector model}) is the counterterm generated by  wave function renormalization of the fields $\phi$ and $s$. Other possible counterterms vanish at the conformal fixed point.

The cubic term in the effective action will be non-local. By conformal invariance, its form is fixed up to an overall constant $\hat Z$, and can be represented diagrammatically as a conformal triangle \cite{Polyakov:1970xd},
\begin{center}
\begin{equation}
\label{definition of vertex Gamma vm}
  \begin{picture}(300,57) (23,35)
    \SetWidth{1.0}
    \SetColor{Black}
    \Line[](80,95)(52,47)
    \Line[](52,47)(108,47)
    \Line[](111,44)(83,96)
    \Vertex(81,97){4}
    \Vertex(52,47){4}
    \Vertex(111,47){4}
    \Text(69,107)[lb]{\scalebox{1}{$s(x_3)$}}
    \Text(20,35)[lb]{\scalebox{1}{$\phi(x_1)$}}
    \Text(115,35)[lb]{\scalebox{01}{$\phi(x_2)$}}
    \Text(-40,66)[lb]{\scalebox{1.1}{$-\frac{2\hat Z \mu^{2\gamma_\phi+\gamma_s}}{\sqrt{N}}~\times$}}
    \Text(102,73)[lb]{\scalebox{0.8}{$2\alpha$}}
    \Text(78,34)[lb]{\scalebox{0.8}{$2\beta$}}
    \Text(50,73)[lb]{\scalebox{0.8}{$2\alpha$}}
    \Text(120,65)[lb]{\scalebox{1.1}{$~=-\frac{2\hat Z}{\sqrt{N}}\,\int\int\int\,
\frac{\mu^{ 2\gamma_\phi+\gamma_s}}
{|x_{13}|^{2\alpha}|x_{32}|^{2\alpha}|x_{12}|^{2\beta}}\,\phi(x_1)\phi(x_2)s(x_3)\,,$}}
  \end{picture}
  \end{equation}
\end{center}
where
\begin{equation}
\alpha = \Delta_\phi-\frac{\gamma_s}{2}\,,\qquad \beta=\Delta_s-\gamma_\phi+\frac{\gamma_s}{2}~.
\end{equation}

Using the above cubic vertex, one can calculate the corresponding conformal three-point function by attaching to it the full $\phi$ and $s$ propagators, which are given by,
\begin{align}
\label{full vm propagators}
\langle \phi(x)\phi(0)\rangle = \frac{C_\phi\,(1+A_\phi)}{|x|^{2(\Delta_\phi+\gamma_\phi)}}\,,\qquad
\langle s(x)s(0)\rangle = \frac{C_s\,(1+A_s)}{|x|^{2(\Delta_s+\gamma_s)}}\,,
\end{align}
where the normalizations are,
\begin{align}
C_\phi = \frac{\Gamma\left(\frac{d}{2}-1\right)}{4\pi^\frac{d}{2}}\,,\qquad
C_s = \frac{2^d\Gamma\left(\frac{d-1}{2}\right)\sin\left(\frac{\pi d}{2}\right)}{\pi^\frac{3}{2}
\Gamma\left(\frac{d}{2}-2\right)}~,
\end{align}
and the anomalous dimensions are \cite{Vasiliev:1981yc,Vasiliev:1981dg,Lang:1993ct,Petkou:1995vu,Petkou:1994ad,Derkachov:1997ch,Leonhardt:2003du,Gracey:2018ame,Alday:2019clp,Giombi:2019upv,Goykhman:2019kcj}, 
\begin{align}
\gamma_\phi = \frac{1}{N}\,\frac{2^d\sin\left(\frac{\pi d}{2}\right)\Gamma\left(\frac{d-1}{2}\right)}
{\pi^\frac{3}{2}(d-2)d\Gamma\left(\frac{d}{2}-2\right)}\,,\qquad
\gamma_s = \frac{1}{N}\,\frac{4\sin\left(\frac{\pi d}{2}\right)\Gamma(d)}
{\pi\Gamma\left(\frac{d}{2}+1\right)\Gamma\left(\frac{d}{2}-1\right)}\,.
\end{align}
Note that the integration vertices $x_1,x_2$ and $x_3$ become conformal and can be integrated over
using the star-triangle relation (\ref{uniqueness scalar}).

We now calculate $\hat Z$, up to  next-to-leading order in the $1/N$ expansion.
To this end, we notice that the full cubic vertex (\ref{definition of vertex Gamma vm}) is determined by the sum of the tree level term in (\ref {action of vector model}), the counterterm, and the quantum loop corrections. To order ${\cal O}(1/N^{\frac{3}{2}})$, it can be represented by the following diagrammatic equation (there is a factor of $-2/\sqrt{N}$ on both sides, which we drop), 
\begin{equation}
\label{Dressed vertex diagram vm}
\hfill\includegraphics[scale=0.85]{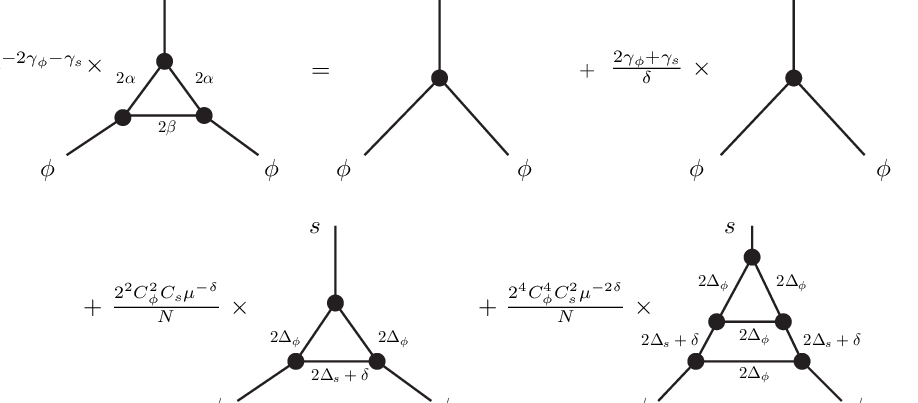}\hspace*{\fill}
\end{equation}

The first term on the right-hand side of (\ref{Dressed vertex diagram vm}) corresponds to the leading order tree level vertex in (\ref {action of vector model}), while the rest of the terms represent ${\cal O}(1/N^{\frac{3}{2}})$ corrections.
Specifically, the last two diagrams account for the loop corrections to the tree level vertex. The scaling dimension of $s$  in these diagrams undergoes a slight shift of $\delta/2$, in order to regularize divergent loops. The last graph on the first line of  (\ref{Dressed vertex diagram vm}) is associated with the counterterm, see  (\ref{action of vector model});  in the limit $\delta\rightarrow 0$, it cancels the divergences of the loop diagrams. The external legs are associated with the fields $\phi$ and $s$, as indicated by the corresponding labels in the figure.

The diagrammatic equation (\ref{Dressed vertex diagram vm}) defines $\hat Z$ up to order ${\cal O}(1/N^{\frac{3}{2}})$. It turns out that the background field method simplifies this equation, and getting an explicit expression for $\hat Z$ boils down to a relatively short calculation.

To begin with, we decompose the Hubbard-Stratonovich field into a constant background $\bar s$ and fluctuating component $s$, \textit{i.e.} we replace $s\rightarrow \bar s+s$. Next, we substitute this decomposition into (\ref{Dressed vertex diagram vm}) and retain only the linear terms in $\bar s$. Now,  replacing the external fields $\phi$ with the full propagators (\ref{full vm propagators}) defines a linear
(in $\bar s$) correction to the two-point function $\langle\phi(x_1)\phi(x_2)\rangle_{\bar s}$, in the presence of a constant background $\bar s$, 
\begin{equation}
\label{Dressed scalar propagator in s background vm}
\hfill\includegraphics[scale=1]{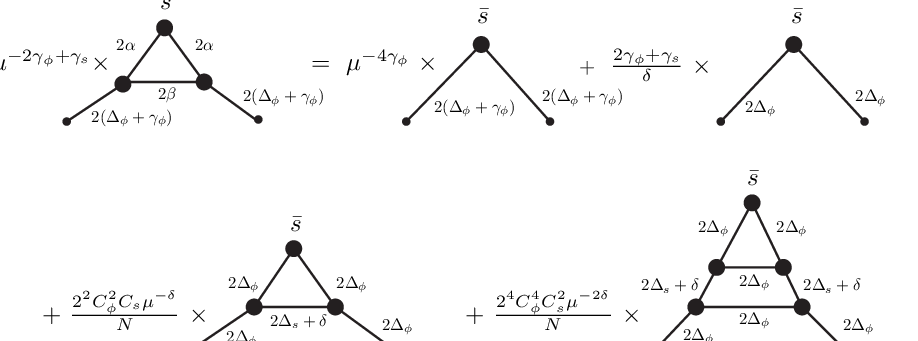}\hspace*{\fill}
\end{equation}
The amplitudes of the full $\phi$-propagators (\ref{full vm propagators}) cancelled out on both sides of this equation.
Furthermore, starting from the integral over the insertion point of the vertex with a constant background $\bar s$, all diagrams in (\ref{Dressed scalar propagator in s background vm}) can be calculated by repeatedly applying the propagator merging relation reviewed in section \ref{sec:useful identities}.

Carrying out this calculation, one gets for the left-hand side,
\begin{align}
\text{ l.h.s. of (\ref{Dressed scalar propagator in s background vm})}&=
\hat Z\,U\left(\frac{d-\gamma_s}{2}-1,\frac{d-\gamma_s}{2}-1,
2+\gamma_s\right)U\left(\frac{d-\gamma_s}{2}-\gamma_\phi,
\frac{d}{2}-1+\gamma_\phi,\frac{\gamma_s}{2}+1\right)\notag\\
&\times U\left(\frac{d}{2}-1+\gamma_\phi,\frac{d-\gamma_s}{2}-1,2+\frac{\gamma_s}{2}-\gamma_\phi\right)
\,\frac{\bar s\,\mu^{-2\gamma_\phi+\gamma_s}}{|x_{12}|^{d-4+2\gamma_\phi-\gamma_s}}\,.
\end{align}
Substituting 
\begin{equation}
\hat Z = Z_0 (1+\delta Z)\,,
\end{equation}
where $\delta Z = {\cal O}(1/N)$,  and expanding in $\gamma_{\phi, s}\sim {\cal O}(1/N)$, yields
\begin{align}
\label{phi phi s triangle}
&\text{ l.h.s. of (\ref{Dressed scalar propagator in s background vm})}=
Z_0\,
\frac{2(d-2)\pi^d\,U\left(\frac{d}{2}-1,\frac{d}{2}-1,2\right)}{(d-4)\Gamma\left(\frac{d}{2}\right)^2(2\gamma_\phi+\gamma_s)}\,
\frac{\bar s}{|x_{12}|^{d-4}}\left(1+\delta Z\right.\\
&+\left.
\frac{((26-3d)d-44)\gamma_s+2((d-6)d+4)\gamma_\phi}{2(d-2)(d-4)}
- (2\gamma_\phi-\gamma_s)\,\log(|x_{12}|\mu)\right)\,.\notag
\end{align}
We wish to equate this with the right-hand side of (\ref{Dressed scalar propagator in s background vm}), which is a  sum of four diagrams which we denote by,
\begin{equation}
\label{phi phi s triangle in terms of sum of vs}
\text{ r.h.s. of (\ref{Dressed scalar propagator in s background vm})}= v_\text{tree} + v_{\textrm{c.t.}}+v_\text{loop}^{(1)}(\delta)+v_\text{loop}^{(2)}\Big(\frac{\delta}{2}\Big)~.
\end{equation}
Let us evaluate each of the four terms.
The tree level term is given by,
\begin{equation}
v_\text{tree}  = U\left(\frac{d}{2}-1+\gamma_\phi,\frac{d}{2}-1+\gamma_\phi,
2-2\gamma_\phi\right)\,\frac{\bar s\,\mu^{-4\gamma_\phi}}{|x_{12}|^{d-4+4\gamma_\phi}}~.
\end{equation}
 This term contributes at leading and subleading order in $1/N$,
\begin{equation}
\label{v0 vm}
v_\text{tree}  = U\left(\frac{d}{2}{-}1,\frac{d}{2}{-}1,2\right)
\,\left(1+\frac{1}{N}\,\frac{2^{d-2} (d-6) \sin \left(\frac{\pi  d}{2}\right) \Gamma \left(\frac{d-1}{2}\right)}{\pi ^{3/2}  \Gamma \left(\frac{d}{2}+1\right)}
-4\gamma_\phi \log(|x_{12}|\mu)\right)
\,\frac{\bar s}{|x_{12}|^{d-4}}\,.
\end{equation}
The counterterm takes the form,
\begin{equation}
\label{vct vm}
v_{\textrm{c.t.}} =U\left(\frac{d}{2}-1,\frac{d}{2}-1,2\right)
\,\frac{2\gamma_\phi+\gamma_s}{\delta}\,\frac{\bar s}{|x_{12}|^{d-4}}~.
\end{equation}
The first loop diagram is given by,
\begin{align}
\label{v1 c vm}
v_\text{loop}^{(1)}(\delta) &= \frac{2^2C_\phi^2 C_s }{N}\,\frac{\bar s}{(\mu |x_{12}|)^{d-4+\delta}}\,
U\left(\frac{d}{2}{-}1,\frac{d}{2}{-}1,2\right)U\left(\frac{d+\delta}{2},\frac{d}{2}{-}1,1{-}\frac{\delta}{2}\right)\\
&\times U\left(\frac{d}{2}{-}1,\frac{d+\delta}{2}{-}1,2{-}\frac{\delta}{2}\right)~. \notag
\end{align}
and the second loop diagram is given by, 
\begin{align}
\label{v2 c vm}
v_\text{loop}^{(2)}\Big({\delta\over 2}\Big)&=  
\frac{2^4C_\phi^4 C_s^2 }{N}\,\frac{\bar s}{(\mu |x_{12}|)^{d-4+2\delta}}\,
U\left(\frac{d}{2}{-}1,\frac{d}{2}{-}1,2\right)U\left(d{-}3,2{+}\frac{\delta}{2},
1{-}\frac{\delta}{2}\right)\\
&\times U\left(\frac{d+\delta}{2}{-}1,2{+}\frac{\delta}{2},\frac{d}{2}{-}1{-}\delta\right)
U\left(\frac{d}{2}{-}1,\frac{d}{2}{+}\delta,1{-}\delta\right)\notag
U\left(\frac{d}{2}{-}1,\frac{d}{2}{-}1{+}\delta,2{-}\delta\right)\,.\notag
\end{align}
Expanding (\ref{v1 c vm}) and (\ref{v2 c vm}) around $\delta = 0$, we obtain, 
\begin{align}
\label{v12 vm}
&v_\text{loop}^{(1)}(\delta) + v_\text{loop}^{(2)}\Big(\frac{\delta}{2}\Big) =U\left(\frac{d}{2}-1,\frac{d}{2}-1,2\right)
\,\frac{\bar s}{|x_{12}|^{d-4}}\\
&\times\,\left(-\frac{2\gamma_\phi+\gamma_s}{\delta}+(2\gamma_\phi+\gamma_s)\,\log(|x_{12}|\mu)
+\frac{1}{N}\,\frac{2^{d-2} (3 (d-6) d+28) \sin \left(\frac{\pi  d}{2}\right) \Gamma \left(\frac{d-1}{2}\right)}{\pi ^{3/2} (d-4) (d-2) \Gamma \left(\frac{d}{2}\right)}\right)\,.\notag
\end{align}
Combining (\ref{v0 vm}), (\ref{vct vm}) and (\ref{v12 vm}), we see that the poles cancel, and we are left with a piece that is finite in the limit $\delta\rightarrow 0$,
\begin{align}
&v_\text{tree} + v_{\textrm{c.t.}}+v_\text{loop}^{(1)}(\delta)+v_\text{loop}^{(2)}\Big(\frac{\delta}{2}\Big) 
\quad \underset{\delta\to 0}{\longrightarrow} \quad
U\left(\frac{d}{2}-1,\frac{d}{2}-1,2\right)\,\frac{\bar s}{|x_{12}|^{d-4}}\label{v total vm}\\
&\times \left(1+\frac{1}{N}\,
\frac{2^{d-3} (d (d (5 d-42)+116)-96) \sin \left(\frac{\pi  d}{2}\right) \Gamma \left(\frac{d-1}{2}\right)}{\pi ^{3/2} (d-4) (d-2) \Gamma \left(\frac{d}{2}+1\right)}-(2\gamma_\phi-\gamma_s)\,\log(|x_{12}|\mu)\right)\,.\notag
\end{align}
Plugging this expression back into (\ref{phi phi s triangle in terms of sum of vs}) and comparing to (\ref{phi phi s triangle}), yields
\begin{align}
\label{Z0 phi4}
Z_0 &= 
-\frac{1}{2\pi^d}\,(2\gamma_\phi+\gamma_s)A(1)^2A\left(\frac{d}{2}-2\right)\Gamma\left(\frac{d}{2}\right)\,,\\
\delta Z &=\frac{1}{N}\,
\frac{2^{d-3} (d (d (5 d-42)+116)-96) \sin \left(\frac{\pi  d}{2}\right) \Gamma \left(\frac{d-1}{2}\right)}{\pi ^{3/2} (d-4) (d-2) \Gamma \left(\frac{d}{2}+1\right)}\notag\\
&=\left(\frac{8}{(d-4)^2}+\frac{2}{d-2}+\frac{6}{d-4}+5\right)\,\gamma_\phi\,.\label{delta Z}
\end{align}
This completes the evaluation of the cubic vertex  (\ref{definition of vertex Gamma vm}). 
This expression matches the result in \cite{Derkachov:1997ch}, which was obtained in a different way.

\section{The $\phi^6$ model in terms of auxiliary fields}
\label{B}

In this appendix we give a slight variant of the derivation of some of the results for the sextic model that were presented in the main body of the paper.
Recall that the action for the $\phi^6$ model is given by (\ref{3d action in terms of aux fields}),
\begin{equation}
S = \int d^3x \left(\frac{1}{2}\,(\partial\phi) ^ 2+ \frac{1}{\sqrt{N}} \, \sigma\,\phi^2 + \frac{g_6}{6\,\sqrt{N}}\,\rho^3 
-\,\sigma\rho\right)~.
\end{equation}
Let us explicitly integrate out the  $\phi$ fields, which results in the effective action,
\begin{equation}
S =\frac{N}{2}\, \textrm{Tr}\log\left(\partial^2 -\frac{2\sigma}{\sqrt{N}}\right)+\int d^3x \left( \frac{g_6}{6\,\sqrt{N}}\,\rho^3 
-\,\sigma\rho\right)\,.
\end{equation}
We now expand the logarithm to  next-to-leading order in $1/N$. Making use of the matrix element,
\begin{equation}
\langle x | (\partial^2)^{-1}\sigma(x)|y\rangle = - \frac{C_\phi}{|x-y|}\,\sigma(y)~, \ \ \ \ \ \  \ C_\phi =\frac{ 1}{4\pi}~,
\end{equation}
and discarding the constant and linear in $\sigma$ terms (which are removed by appropriate counterterms) we obtain,
\begin{align}
S &=\int d^3x \left( \frac{g_6}{6\,\sqrt{N}}\,\rho^3 -\sigma\rho\right)
-\frac{1}{16\pi^2}\int d^3x_1 d^3 x_2\frac{\sigma(x_1)\sigma(x_2)}
{|x_{12}|^2}\\
&+\frac{1}{\sqrt{N}}\frac{1}{48\pi^3}\int d^3x_1 d^3 x_2 d^3 x_3
\frac{\sigma(x_1)\sigma(x_2)\sigma(x_3)}{|x_{12}||x_{13}||x_{23}|}+{\cal O}(1/N^{\frac{3}{2}})\,,\notag
\end{align}
In order to diagonalize the quadratic part, we define a new field $s$ in terms of $\sigma$, 
\begin{equation}
\label{sigma redefinition}
\sigma(x) = s (x) + \frac{4}{\pi^2} \,\int d^3x'\,\frac{\rho(x')}{|x-x'|^4}~.
\end{equation}
Using the star-triangle relation, as well as the  inverse propagator relation,
\begin{equation}
\label{inverse propagator}
\int d^3 x_2\,\frac{1}{|x_{12}|^4|x_{23}|^2} = -2\pi^4\,\delta^{(3)}(x_{13})~,
\end{equation}
we obtain the effective action, up to next-to-leading order, 
\begin{align}
\label{action after integrating out phi}
S &=\! -\! \int\! d^3x_1 d^3 x_2 \left(\frac{2}{\pi^2}\,\frac{\rho(x_1)\rho(x_2)}{|x_{12}|^4}
+\frac{1}{16\pi^2}\,\frac{s(x_1)s(x_2)}{|x_{12}|^2}\right) \notag \\
&+\frac{g_6}{6\sqrt{N}}\int\! d^3x_1\, \rho(x_1)^3 +\!\frac{1}{\sqrt{N}}\frac{4}{\pi^2}\int d^3x_1 d^3x_2 \frac{\rho(x_1)^2s(x_2)}{|x_{12}|^2} \notag\\
&+\frac{1}{\sqrt{N}}\frac{4}{3\pi^9}\,\int 
 \frac{\prod_{i=1}^3 d^3 x_i ~d^3 x'_i}{|x_1-x'_1|^4|x_2-x'_2|^4|x_3-x'_3|^4}
\frac{\rho(x_1)\rho(x_2)\rho(x_3)}{|x'_{12}|x'_{13}||x'_{23}|}
\notag\\
&+\frac{1}{\sqrt{N}}
\int \prod_{i=1}^3 d^3 x_i \( \frac{1}{48\pi^3}\frac{s(x_1)s(x_2)s(x_3)}{|x_{12}| |x_{13}| |x_{23}|}    -\frac{1}{2\pi^4}\frac{\rho(x_1)s(x_2)s(x_3)}{|x_{12}|^2 |x_{13}|^2}  \) ~.
\end{align}
Inverting the bilinear part of the action (\ref{action after integrating out phi}) through the use of (\ref{inverse propagator}), we obtain the leading order propagators
\begin{align}
\label{rho and sigma two point functions}
\langle \rho(x)\rho(y)\rangle = \frac{1}{8\pi^2}\,\frac{1}{|x-y|^2}\,,\qquad
\langle s(x)s(y)\rangle = \frac{4}{\pi^2}\,\frac{1}{|x-y|^4}\,.
\end{align}

The Feynman rules for the amputated cubic vertices are\footnote{It is assumed that all of the three $s$
fields are located at different points $x_i$, $i=1,2,3$, and are connected with the $\phi$-propagator lines. If the locations of any of the two $s$ fields coincide, the corresponding Feynman rule vanishes, owing to the conformal relation $\langle \phi^2\rangle =0$.}
\begin{center}
  \begin{picture}(380,106) (39,-15)
    \SetWidth{1.0}
    \SetColor{Black}
    \scalebox{0.7}{
    \Line[](40,42)(112,42)
    \Vertex(112,42){4}
    \Line[](112,42)(160,90)
    \Line[](112,42)(160,-6)
    \Text(184,32)[lb]{\scalebox{1.4}{$= -\frac{g_6}{\sqrt{N}}$}}
    \Text(154,75)[lb]{\scalebox{1.2}{$\rho$}}
    \Text(154,5)[lb]{\scalebox{1.2}{$\rho$}}
    \Text(40,30)[lb]{\scalebox{1.2}{$\rho$}}
    }
    \scalebox{0.8}{
    \Line[](258,42)(340,42)
    \Vertex(340,42){4}
    \Line[](340,42)(412,42)
    \Text(375,50)[lb]{\scalebox{1.2}{$2$}}
    \Text(425,70)[lb]{\scalebox{1.2}{$\rho$}}
    \Text(425,5)[lb]{\scalebox{1.2}{$\rho$}}
    \Line[](412,42)(460,90)
    \Line[](412,42)(460,-14)
    \Vertex(412,42){4}
    \Text(480,34)[lb]{\scalebox{1.4}{$=-\frac{8}{\pi^2\sqrt{N}}$}}
    \Text(300,50)[lb]{\scalebox{1.2}{$s$}}
    }
  \end{picture}
\end{center}
\begin{center}
  \begin{picture}(440,93) (60,0)
    \SetWidth{1.0}
    \SetColor{Black}
    \scalebox{0.6}{
    \Line[](190,144)(190,79)
    \Text(195,140)[lb]{\scalebox{1.2}{$\rho$}}
    \Line[](190,79)(154,36)
    \Line[](190,79)(222,36)
    \Line[](155,36)(222,36)
    \Vertex(190,110){4}
    \Text(195,95)[lb]{\scalebox{1.2}{$4$}}
    \Vertex(190,79){4}
    \Vertex(155,36){4}
    \Vertex(127,24){4}
    \Text(133,34)[lb]{\scalebox{1.2}{$4$}}
    \Vertex(250,24){4}
    \Text(240,34)[lb]{\scalebox{1.2}{$4$}}
    \Vertex(222,36){4}
    \Line[](222,36)(281,11)
    \Line[](155,36)(97,11)
    \Text(260,64)[lb]{\scalebox{1.8}{$=-\frac{8}{\pi^9\sqrt{N}}$}}
    \Text(160,58)[lb]{\scalebox{1.2}{$1$}}
    \Text(215,58)[lb]{\scalebox{1.2}{$1$}}
    \Text(188,21)[lb]{\scalebox{1.2}{$1$}}
    \Text(100,0)[lb]{\scalebox{1.2}{$\rho$}}
    \Text(270,0)[lb]{\scalebox{1.2}{$\rho$}}
   \Line[](440,144)(440,79)
    \Text(445,110)[lb]{\scalebox{1.2}{$s$}}
    \Line[](440,79)(404,36)
    \Line[](440,79)(472,36)
    \Line[](405,36)(472,36)
    \Vertex(440,79){4}
    \Vertex(405,36){4}
    \Vertex(472,36){4}
    \Line[](472,36)(531,11)
    \Line[](405,36)(347,11)
    \Text(510,64)[lb]{\scalebox{1.8}{$=-\frac{1}{8\pi^3\sqrt{N}}$}}
    \Text(410,58)[lb]{\scalebox{1.2}{$1$}}
    \Text(465,58)[lb]{\scalebox{1.2}{$1$}}
    \Text(438,21)[lb]{\scalebox{1.2}{$1$}}
    \Text(370,10)[lb]{\scalebox{1.2}{$s$}}
    \Text(500,10)[lb]{\scalebox{1.2}{$s$}}
    \Line[](690,144)(690,79)
    \Text(695,110)[lb]{\scalebox{1.2}{$\rho$}}
    \Line[](690,79)(654,36)
    \Line[](690,79)(722,36)
    \Vertex(690,79){4}
    \Vertex(655,36){4}
    \Vertex(722,36){4}
    \Line[](722,36)(781,11)
    \Line[](655,36)(597,11)
    \Text(750,64)[lb]{\scalebox{1.8}{$=\frac{1}{\pi^4\sqrt{N}}$}}
    \Text(660,58)[lb]{\scalebox{1.2}{$2$}}
    \Text(715,58)[lb]{\scalebox{1.2}{$2$}}
    \Text(620,10)[lb]{\scalebox{1.2}{$s$}}
    \Text(750,10)[lb]{\scalebox{1.2}{$s$}}
   }
  \end{picture}
\end{center}
When all three fields $\rho$ are located at different points, we can combine two $\mathcal{O}(\rho^3)$ vertices into one,
with the total amplitude $\frac{64-g_6}{\sqrt{N}}$. In particular, we recover the leading order three-point function (\ref{leading normalized rho rho rho}).

Using these rules, yields
\begin{align}
\langle s(x_1)s(x_2)\rho(x_3)\rangle &= {\cal O}\left(\frac{1}{N^{3/2}}\right)\,,\\
\langle s(x_1)s(x_2)s(x_3)\rangle &=  {\cal O}\left(\frac{1}{N^{3/2}}\right)\,,\\
\langle \rho(x_1)\rho(x_2)s(x_3)\rangle&=\frac{4}{\pi\sqrt{N}}\,
\frac{1}{|x_{13}|^2|x_{23}|^2}+{\cal O}\left(\frac{1}{N^{3/2}}\right)\,.
\end{align}
where we used (\ref{inverse propagator}) and the star-triangle relation, and $s\rightarrow \sqrt{C_s}\,s$.

We conclude this appendix by reproducing the $\rho^3$
effective vertex at the next-to-leading order. The contributing diagrams are
\begin{center}
  \begin{picture}(400,120) (10,-17)
    \SetWidth{1.0}
    \SetColor{Black}
    \scalebox{0.6}{
    \Line[](114,122)(114,80)
    \Line[](114,80)(54,2)
    \Line[](114,80)(174,2)
    \Line[](54,2)(174,2)
    \Line[](54,2)(12,-16)
    \Line[](174,2)(216,-16)
    \Text(-10,34)[lb]{\scalebox{1.2}{$W_1 =$}}
    \Vertex(54,2){4}
    \Vertex(174,2){4}
    \Vertex(114,80){4}
    \Text(60,44)[lb]{\scalebox{1.2}{$2\Delta_\rho$}}
    \Text(150,44)[lb]{\scalebox{1.2}{$2\Delta_\rho$}}
    \Text(109,-14)[lb]{\scalebox{1.2}{$2\Delta_\rho$}}
    \Line[](404,122)(404,80)
    \Line[](404,80)(344,2)
    \Line[](404,80)(464,2)
    \Line[](344,2)(464,2)
    \Line[](344,2)(302,-16)
    \Line[](464,2)(506,-16)
    \Text(280,34)[lb]{\scalebox{1.2}{$W_2 =$}}
    \Vertex(344,2){4}
    \Vertex(464,2){4}
    \Vertex(404,80){4}
    \Vertex(386,56){4}
    \Vertex(422,56){4}
    \Vertex(362,26){4}
    \Vertex(446,26){4}
    \Vertex(380,2){4}
    \Vertex(428,2){4}
    \Text(382,68)[lb]{\scalebox{1.2}{$2$}}
    \Text(420,68)[lb]{\scalebox{1.2}{$2$}}
    \Text(352,44)[lb]{\scalebox{1.2}{$2\Delta_s$}}
    \Text(440,44)[lb]{\scalebox{1.2}{$2\Delta_s$}}
    \Text(342,14)[lb]{\scalebox{1.2}{$2$}}
    \Text(360,-10)[lb]{\scalebox{1.2}{$2$}}
    \Text(399,-14)[lb]{\scalebox{1.2}{$2\Delta_s$}}
    \Text(447,-10)[lb]{\scalebox{1.2}{$2$}}
    \Text(461,14)[lb]{\scalebox{1.2}{$2$}}
   }
  \end{picture}
\end{center}
\begin{center}
  \begin{picture}(400,120) (10,-17)
    \SetWidth{1.0}
    \SetColor{Black}
    \scalebox{0.6}{
    \Line[](114,122)(114,80)
    \Line[](114,80)(54,2)
    \Line[](114,80)(174,2)
    \Line[](54,2)(174,2)
    \Line[](54,2)(12,-16)
    \Line[](174,2)(216,-16)
    \Text(-10,34)[lb]{\scalebox{1.2}{$W_3 =$}}
    \Vertex(54,2){4}
    \Vertex(174,2){4}
    \Vertex(114,80){4}
    \Vertex(96,56){4}
    \Vertex(132,56){4}
    \Vertex(72,26){4}
    \Vertex(156,26){4}
    \Text(92,68)[lb]{\scalebox{1.2}{$2$}}
    \Text(130,68)[lb]{\scalebox{1.2}{$2$}}
    \Text(60,44)[lb]{\scalebox{1.2}{$2\Delta_s$}}
    \Text(150,44)[lb]{\scalebox{1.2}{$2\Delta_s$}}
    \Text(51,14)[lb]{\scalebox{1.2}{$2$}}
    \Text(109,-15)[lb]{\scalebox{1.2}{$2\Delta_\rho$}}
    \Text(171,14)[lb]{\scalebox{1.2}{$2$}}
    \Line[](404,122)(404,80)
    \Line[](404,80)(344,2)
    \Line[](404,80)(464,2)
    \Line[](344,2)(464,2)
    \Line[](344,2)(302,-16)
    \Line[](464,2)(506,-16)
    \Text(280,34)[lb]{\scalebox{1.2}{$W_4 =$}}
    \Vertex(344,2){4}
    \Vertex(464,2){4}
    \Vertex(404,80){4}
    \Vertex(380,2){4}
    \Vertex(428,2){4}
    \Text(350,44)[lb]{\scalebox{1.2}{$2\Delta_\rho$}}
    \Text(440,44)[lb]{\scalebox{1.2}{$2\Delta_\rho$}}
    \Text(360,-10)[lb]{\scalebox{1.2}{$2$}}
    \Text(399,-15)[lb]{\scalebox{1.2}{$2\Delta_s$}}
    \Text(447,-10)[lb]{\scalebox{1.2}{$2$}}
    }
  \end{picture}
\end{center}
Using   (\ref{inverse propagator}), we obtain
\begin{align}
W_1 &=  C_\rho^3 \left(\frac{64-g_6}{\sqrt{N}}\right)^3\,W_0\,,\quad
W_2 = C_s^3 \left(\frac{1}{\pi^4\sqrt{N}}\right)^3\,(-2\pi^4)^3 \,W_0\,,\\
W_3 &{=} 3C_\rho C_s^2\frac{1}{\pi^4\sqrt{N}} \left(-\frac{8}{\pi^2\sqrt{N}}\right)^2\, ({-}2\pi^4)^2\, W_0\,,\;\;
W_4 {=} 3C_\rho^2 C_s\frac{64{-}g_6}{\sqrt{N}} \left({-}\frac{8}{\pi^2\sqrt{N}}\right)^2\,({-}2\pi^4)\,W_0\,,\notag
\end{align}
where $W_0 = \int d^3x_{1,2,3}\,\frac{\rho(x_1)\rho(x_2)\rho(x_3)}{(|x_{12}||x_{13}||x_{23}|)^2}$.
The cubic vertex is then given by
\begin{equation}
\sum_{i=1}^4 W_i = \frac{g_6^2(192-g_6)}{512\pi^6}\,\,\int d^3x_{1,2,3}\,\frac{\rho(x_1)\rho(x_2)\rho(x_3)}{(|x_{12}||x_{13}||x_{23}|)^2}\,.
\end{equation}
in agreement with (\ref{V3 1 over N}).\footnote{Recall that effective vertex and the corresponding diagram
are related by the symmetry factor of $-1/3!$} 
This vertex vanishes at the fixed point  $g_6^\star = 192$.

\end{document}